\documentclass[12pt,preprint]{aastex}
\usepackage{epsf,color,cancel,algorithmic,amsmath,subfigure}

\newcommand{\sfrac}[2]{\mathchoice
  {\kern0em\raise.5ex\hbox{\the\scriptfont0 #1}\kern-.15em/
   \kern-.15em\lower.25ex\hbox{\the\scriptfont0 #2}}
  {\kern0em\raise.5ex\hbox{\the\scriptfont0 #1}\kern-.15em/
   \kern-.15em\lower.25ex\hbox{\the\scriptfont0 #2}}
  {\kern0em\raise.5ex\hbox{\the\scriptscriptfont0 #1}\kern-.2em/
   \kern-.15em\lower.25ex\hbox{\the\scriptscriptfont0 #2}}
  {#1\!/#2}}

\newcommand{\myhalf}{\sfrac{1}{2}}

\newcommand{\eb}{\boldsymbol{e}}
\newcommand{\Ub}{\boldsymbol{U}}

\newcommand{\nablab}{\mathbf{\nabla}}

\newcommand{\dt}{\Delta t}

\newcommand{\omegadot}{\dot{\omega}}

\newcommand{\Hnuc}{H_{\rm nuc}}
\newcommand{\kth}{k_{\rm th}}

\newcommand{\Gammaonebar}{\overline{\Gamma}_1}
\newcommand{\Sbar}{\overline{S}}

\newcommand{\He}{$^4$He}
\newcommand{\C}{$^{12}$C}
\newcommand{\Fe}{$^{56}$Fe}

\newcommand{\avgtwod}[1]{\langle #1 \rangle}

\newcommand{\mymax}[1]{\left(#1\right)_{\rm max}}

\newcommand{\cold}{{\tt cold}}
\newcommand{\hot}{{\tt hot}}

\begin{document}
%==========================================================================
% Title
%==========================================================================
\title{Multidimensional Modeling of Type I X-ray
  Bursts. I. Two-Dimensional Convection Prior to the Outburst of a
  Pure \He\ Accretor} 
\shorttitle{Multidimensional Modeling of Type I X-ray Bursts I.}
\shortauthors{Malone et al.}

\author{C.~M.~Malone\altaffilmark{1},
        A.~Nonaka\altaffilmark{2},
        A.~S.~Almgren\altaffilmark{2},
        J.~B.~Bell\altaffilmark{2},
        M.~Zingale\altaffilmark{1}}
\email{cmalone@mail.astro.sunysb.edu}

\altaffiltext{1}{Dept. of Physics \& Astronomy,
                 Stony Brook University,
		 Stony Brook, NY 11794-3800}

\altaffiltext{2}{Center for Computational Sciences and Engineering,
                 Lawrence Berkeley National Laboratory,
                 Berkeley, CA 94720}

%==========================================================================
% Abstract and Keywords
%==========================================================================
\begin{abstract}
  We present multidimensional simulations of the early convective
  phase preceding ignition in a Type I X-ray burst using the low Mach
  number hydrodynamics code, {\tt MAESTRO}.  A low Mach number
  approach is necessary in order to perform long-time integration
  required to study such phenomena.  Using {\tt MAESTRO}, we are able
  to capture the expansion of the atmosphere due to large-scale
  heating while capturing local compressibility effects such as those
  due to reactions and thermal diffusion.  We also discuss the
  preparation of one-dimensional initial models and the subsequent
  mapping into our multidimensional framework.  Our method of initial
  model generation differs from that used in previous multidimensional
  studies, which evolved a system through multiple bursts in
  one dimension before mapping onto a multidimensional grid.  In our
  multidimensional simulations, we find that the resolution necessary
  to properly resolve the burning layer is an order of magnitude
  greater than that used in the earlier studies mentioned above.  We
  characterize the convective patterns that form and discuss their
  resulting influence on the state of the convective region, which is
  important in modeling the outburst itself.
\end{abstract}
\keywords{convection---hydrodynamics---methods: numerical---stars: neutron---X-rays: bursts}

%==========================================================================
% Introduction
%==========================================================================
\section{Introduction}\label{Sec:Introduction}
Type I X-ray bursts are possibly the most frequent thermonuclear
explosions in the universe and provide a large amount of observational
data that can be used to determine the properties of matter near the
surface of a neutron star.  To make meaningful inferences about these
properties from observational data, however, we must have a proper
theoretical understanding of the bursting phenomena \citep{BHATTA10}.
The basic XRB paradigm takes place in a mass-transferring, low-mass
X-ray binary (LMXB) system in which the neutron star's companion has
filled its Roche lobe and is dumping H- and/or He-rich material onto
the surface of the neutron star.  Depending on the accretion rate and
composition, there are several burning regimes that will trigger an
XRB (see \cite{BILDSTEN00} for an overview).  The general idea is that
a column of accreted material---or heavier-element ash from prior
stable burning of accreted material---builds up until the
temperature sensitivity of the energy generation rate at the base of
the layer exceeds that of the local cooling rate and a thin-shell
thermal instability forms.  The instability eventually causes a
runaway of unstable burning resulting in an outburst.

One-dimensional hydrodynamic studies reproduce many of the observable
features of XRBs such as burst energies ($\sim 10^{39}$ erg), rise
times (seconds), durations ($10$'s -- $100$'s of seconds) and
recurrence times (hours to days)
(\citealp{WOO_WEAV84,TAAM_ETAL93,HEG_ETAL07}; also see
\cite{STRO_BILD06} for a review of XRBs).  By construction, however,
one-dimensional models assume that the fuel is burned uniformly over
the surface of the star, which is highly unlikely given the large
disparity between the thermalization and burning timescales of the
accreted material \citep{SHARA82}.  Furthermore, the {\em Rossi X-ray
  Timing Explorer} satellite has observed coherent oscillations in the
lightcurves of $\sim 20$ outbursts from LMXB systems (first by
\citealp{STRO_ETAL96}; more recently by \citealp{ALTAMIRANO_ETAL10}
and references therein).  The asymptotic evolution of the frequency of
such oscillations suggests they are modulated by the neutron star spin
frequency \citep{MUNO_ETAL02}.  Oscillations observed during the
rising portion of an outburst lightcurve are therefore indicative of a
spreading burning front being brought in and out of view by stellar
rotation.  Additionally, oscillations observed during the decay phase
of the burst are thought to be caused by unstable surface modes that
may depend critically on the local heating and cooling rates during
the burst \cite[and references therein]{NAR_COOP07}.  The manner in
which the burning front spreads and propagates throughout the accreted
atmosphere is not well known, and a proper multidimensional modeling
of the conditions in the atmosphere prior to outburst is needed
\citep[e.g.][]{FRY_WOOS_PROPAGATION_82}.

Prior to the actual outburst, the burning at the base of the ignition
column drives convection throughout the overlying layers and
determines the state of the material in which the burning front will
propagate.  One-dimensional simulations of XRBs usually attempt to
parameterize the convective overturn and mixing using astrophysical
mixing-length theory \citep{BOHM-VITENSE58} or through various
diffusive processes (see \citealt{HEGER_ETAL00} for a thorough
discussion).  Recent multidimensional simulations of stellar
convection \cite[see][and references therein]{ARNETT_ETAL09}, however,
show a large discrepancy in, for example, the velocity of a typical
convective eddy when compared to one-dimensional models in the case of
stellar evolution codes that use mixing-length theory.  Indeed, there
has recently been an effort put forth in the astrophysical community,
the so-called Convection Algorithms Based on Simulations, or CABS, to
derive from multidimensional simulations a more physically motivated
prescription for handling convection in one dimension
\citep{ARNETT_ETAL08}.  To date, such methods have not propagated into
the XRB-simulation community and a proper treatment of convection,
without assumptions, requires simulation in multiple dimensions.

Multidimensional simulations of any aspect of XRBs, however, have
hitherto been rather restrictive.  A burning front can propagate
either supersonically as a detonation or subsonically as a
deflagration.  Full hydrodynamic XRB detonation models in the spirit
of \cite{FRY_WOOS_DETONATION_82} or \cite{ZINGALE_ETAL01} require a
thick ($\sim 100$ m) accreted helium layer.  Such deep layers are only
produced by very low accretion rates, which are inconsistent with the
majority of rates inferred from observations of XRBs, and therefore
the burning front in most XRBs likely propagates as a deflagration.
Deflagration models are difficult to compute with standard
compressible hydrodynamics codes due to the long integration times
required.  One possible solution is to eliminate the effect of
acoustic waves in the system, allowing the time step to be controlled
by the fluid velocity, rather than the sound speed.  Such a method can
be derived using low Mach number asymptotics; classic examples of low
Mach number approaches include the incompressible, anelastic
\citep{OguPhi62} and Boussinesq \citep{Bou03} approximations.  To this
end, \cite{SPIT_ETAL02} used a simple, shallow-water, 2-layer,
incompressible fluid to model the vertical structure of a deflagration
front and showed how rotation coupled with convection may play an
important role in regulating the spread of the front over the surface
of the neutron star.

More recently, \cite{Lin:2006} developed and applied a low Mach number
approximation method to the problem of convective burning at the base
of an accreted layer in an XRB system.  Their method, however, was
first order accurate in space and time and did not allow for the
evolution of the hydrostatic base state, a feature that is needed to
capture the expansion of the atmosphere in response to heating.
Furthermore, Lin et al. did not model the surface of the accreted
layer, which is vital to understanding bursts that exhibit
photospheric radius expansion (PRE bursts); such bursts are crucial in
determining the stellar properties of neutron stars \citep[and
  references therein]{STEINER_ETAL10}.

In this study we use {\tt MAESTRO} \citep{MAESTRO:Multilevel}, a
multidimensional low Mach number hydrodynamics algorithm for
astrophysical flows, to model the convection leading up to an outburst
of a pure \He\ accretor.  {\tt MAESTRO} is a second-order accurate,
conservative method that uses rectangular grid cells.  The algorithm
is capable of capturing the expansion of the atmosphere due to
large-scale heating, while capturing local compressibility effects due
to reactions, thermal diffusion and compositional changes.  A
semi-analytic method is used to generate one-dimensional initial
models.  These models are then augmented by being evolved in a
one-dimensional stellar evolution code; this evolution allows for the
approximation of the convective cooling and leads to a model that is
closer to satisfying the thin shell instability condition.  The
resulting model is then mapped into {\tt MAESTRO} and evolved in
multiple dimensions.

The main goals of this paper are to explore and describe the
challenges of modeling XRBs in multiple dimensions and to better
understand the convective phase that precedes the outburst.  The
remainder of this paper is as follows.  In Section \ref{Sec:Initial
  Models}, we describe the generation of the initial models and the
subsequent mapping into a multidimensional framework.  In Section
\ref{Sec:Hydrodynamics Algorithm}, we describe the {\tt MAESTRO}
algorithm, including the addition of two new modules not present in
the original algorithm.  Specifically, we have added thermal
conduction and a ``volume discrepancy'' correction term to the
velocity field to ensure that the solution does not diverge from the
equation of state.  In Section \ref{Sec:Results}, we describe the
results of our multidimensional simulations.  In particular we discuss
the resolution needed to properly resolve the burning layer, the
effects of including the thermal diffusion and volume discrepancy
correction terms, the expansion of the base state due to heating and
finally the nature of the convective behavior and its effect on the
atmosphere.  We conclude in Section \ref{Sec:Conclusions} by
summarizing our results and describing our plans to extend the current
study to mixed H/He XRB sources.

%==========================================================================
% Initial Models
%==========================================================================
\section{Initial Models}\label{Sec:Initial Models}
We begin our calculations by generating a one-dimensional initial
model of the accreted layer in hydrostatic (HSE) and thermal
equilibria on the surface of a neutron star.  We assume a
plane-parallel geometry---that is, the gravitational acceleration,
$g$, is assumed constant throughout the domain, which is justified
because the thickness of the accreted layer ($\sim 10$ m) is much less
than the radius of the neutron star ($\sim 10$ km).  We assume a
\He\ layer is accreted on top of a \Fe\ neutron star with a trace
abundance ($10^{-10}$) of \C.  We choose a pure \He\ accretor both
because the corresponding nuclear reaction network, $3\alpha$ burning,
is simple compared to the slow, $\beta$-decay-limited burning
processes in bursts involving H, and because ultra-compact XRB sources
are possible pure \He\ accretors (4U 1820-30, for example;
\citealt{CUMMING_03}).  We include the forward and reverse $3\alpha$
reaction rates as given in \cite{caughlan-fowler:1988} with electron
screening contributions from \cite{Graboske_etal:1973} for the weak
regime and from \cite{Alastuey_Jancovici:1978} for the strong regime.

There are several approaches to one-dimensional model generation in
the literature.  In our approach, we begin with a semi-analytic
initial model and then augment the model to account for convective
cooling.  We also discuss proper mapping of the one-dimensional model
to our multidimensional framework.

\subsection{Semi-Analytic Models}\label{Sec:Semi-analytic Models}
The semi-analytic approach to model generation involves integration
of the heat equation and an entropy equation,
\begin{eqnarray}
  \frac{dT}{dy} &=& \frac{3\kappa F}{4acT^3}\label{eq:heat}\\
  \frac{dF}{dy} &=& 0\label{eq:entropy},
\end{eqnarray}
where $c$ is the speed of light, $a$ the radiation constant, $\kappa$
the opacity (including radiative and conductive contributions), $T$
the temperature, $F$ the outward heat flux and $dy = -\rho dr$ with
$y(r)$ the column-depth (see \citealt{CUMMING_BILDSTEN_00} for details
of this method).  Note that (\ref{eq:heat}) can give a thermal profile
that is superadiabatic---in practice, the thermal gradient is
restricted to be $dT/dy \leq \left(dT/dy\right)_s$ where the subscript
$s$ means along an adiabat.  Also note that for simplicity, equation
(\ref{eq:entropy}) neglects any compressional heating contributions
from the accretion itself and assumes the accreted material is not
burning during the accretion phase---this is a steady-state
configuration.  There is, however, an outward heat flux from
pycnonuclear reactions deep within the neutron star crust; we
approximate this flux as a constant value throughout the accreted
layer, $F = 200$ keV per nucleon.  The integration starts at the top
of the \He\ atmosphere (arbitrarily at $y_\text{top}=10^3$ g
cm$^{-2}$) where a radiative zero solution is assumed, and continues
until the thin shell instability condition \citep{FL87_XRB},
\begin{equation}\label{eq:thermal instability}
  \frac{d\epsilon_{3\alpha}}{dT} > \frac{d\epsilon_\text{cool}}{dT},
\end{equation}
is reached at $y=y_\text{base}$.  The local cooling rate is typically
approximated from (\ref{eq:heat}) and (\ref{eq:entropy}) as
\begin{equation}\label{eq:approx cooling}
\epsilon_\text{cool} \approx \frac{ac T^4}{3\kappa y^2}.
\end{equation}
When (\ref{eq:thermal instability}) is attained, the composition for
$y > y_\text{base}$ is switched to \Fe\  and integration of
(\ref{eq:heat}) and (\ref{eq:entropy}) resumes until a thick enough
substrate is formed such that $y_\text{base}$ is sufficiently far from
the bottom of the computational domain, $y(r=0)=10^{12}$ g cm$^{-2}$ in
our studies.  

The approximation, (\ref{eq:approx cooling}), works well in one
dimension because the only efficient way the system can cool
(neglecting weak reactions) is via conduction and radiation, which
enter through the opacity.  When more spatial dimensions are added to
the system and there is heating from below from nuclear reactions, the
fluid is free to overturn and cool via convection.  Now we have a
situation where the local multidimensional cooling rate,
$\epsilon_\text{cool, multi-d} = \epsilon_\text{cool} +
\epsilon_\text{conv}$, exceeds the initial approximation and
(\ref{eq:thermal instability}) may no longer be satisfied.  Therefore,
such a semi-analytic model is no longer close to runaway and to evolve
the system in multiple dimensions until (\ref{eq:thermal instability})
is reached is intractable even with the advantages of a low Mach
number approximation code.

\subsection{{\tt Kepler}-supplemented Models}\label{Sec:Kepler-supplemented Models}
One way to overcome the difficulties with evolving the model described
in the previous section in multiple dimensions is to explicitly
include an {\it effective} convective cooling term in the
approximation to the local cooling given by equation (\ref{eq:approx
  cooling}).  This {\it effective} convective cooling can be included
via mixing-length theory typically found in stellar evolution codes.
Using the semi-analytic model described above as initial conditions,
the one-dimensional stellar evolution code, {\tt Kepler}
\citep{Kepler}, was used to construct the remainder of the underlying
neutron star with $R_\text{ns} = 10$ km and $M_\text{ns} = 1.87
M_\odot$ (Woosley, 2010; private communication).  The system is then
allowed to evolve in one dimension whereupon nuclear burning heats the
base of the layer, and the convection prescription develops a
well-mixed and nearly adiabatic region of \C\ ash overlying the
\He\ base.  This results in a model that is much closer to satisfying
the thermal instability criterion, (\ref{eq:thermal instability}),
when mapped into multiple dimensions.

\subsection{Mapping to Multiple Dimensions}
The data from {\tt Kepler} are given in a Lagrangian (mass) coordinate
system and we need to convert them to an Eulerian (physical)
coordinate system for use in {\tt MAESTRO}.  We use a procedure
similar to that found in \cite{ppm-hse} to ensure our initial model is
in HSE.  Given the density, temperature and composition from the {\tt
  Kepler} evolution, we call the equation of state to get the
pressure.  We then discretize the HSE equation and solve for the
non-uniform Eulerian grid spacing corresponding to the Lagrangian grid
points,
\begin{equation}\label{eq:hse}
   r_i = r_{i-1} - \frac{1}{g}\frac{p_i - p_{i-1}}{\myhalf
     \left(\rho_i + \rho_{i-1}\right)},
\end{equation}
where $r$ is the radial coordinate, $p$ the pressure and $\rho$ the
density.  We set $r_1=0$ to complete the description of the grid.  The
transition from the pure \Fe\ neutron star (at $r_\text{trans}$) to
the \He\ atmosphere (at $r_{\text{trans}+1}$) is a step function as a
result of the initial Lagrangian data.  Such sharp transitions can be
a source of numerical noise and oscillations as the solution evolves
on an Eulerian grid.  To minimize the numerical noise, we smooth the
interface by adding $n$ uniformly distributed coordinate points
between $r_\text{trans}$ and $r_{\text{trans}+1}$.  The temperature at
these new points is linearly interpolated between $T_\text{trans}$ and
$T_{\text{trans}+1}$.  Then $X(\text{\He})$ and $X(\text{\C})$ at the
new points are filled with a $\tanh$ profile:
\begin{equation}\label{eq:tanh profile}
  \phi_i = \alpha\tanh\left(\frac{r_i - r_\text{c}}{\varphi}\right) +
  \phi_\text{c}
\end{equation}
where $\alpha
=\left(\phi_{\text{trans}+1}-\phi_\text{trans}\right)/2$,
$r_\text{c}=\left(r_\text{trans} + r_{\text{trans}+1}\right)/2$,
$\phi_\text{c} =
\left(\phi_\text{trans}+\phi_{\text{trans}+1}\right)/2$ and $\varphi$
is a parameter to set the smoothness.  $X(\text{\Fe})$ is then found
from the constraint $\sum_k X_k = 1$, and $p$ and $\rho$ are found by
using an iterative Newton-Raphson technique with the equation of state
and (\ref{eq:hse}) at these new points.  This smoothed model is then
linearly interpolated onto a completely uniform grid, with $r_i =
r_{i-1} + \Delta r$, and is again put into HSE using (\ref{eq:hse})
and the equation of state.  Values of $n=50$ and $\varphi = 3$ were
used to smooth the models presented in this work.

Figure \ref{Fig:stan model} shows the result of this procedure for two
models that were evolved in {\tt Kepler} until the base of the
\He\ atmosphere had reached a temperature of $3.67 \times 10^8$ K
(solid line, hereafter referred to as the \cold\ model) and $5.39
\times 10^8$ K (dotted line, hereafter referred to as the
\hot\ model).  The density at the base of the \He\ layer for the
\cold\ model is $1.4 \times 10^6$ g cm$^{-3}$ and is $1.2 \times
10^{6}$ g cm$^{-3}$ for the \hot\ model.  For comparison, the initial
model of \cite{Lin:2006} had a base temperature and density of
$2\times 10^8$ K and $4\times 10^{6}$ g cm$^{-3}$, respectively.  The
\cold\ model has a peak in \C\ production around $r = 382$ cm (i.e.,
the base of the \He\ layer in both models) that appears smoothed in
the more evolved \hot\ model.  Both models, however, have an extended
region of well-mixed \C\ that extends to $r = 624$ cm ($r = 812$ cm)
for the \cold\ (\hot) model.  These initial models contain no
multidimensional velocity information from the {\tt Kepler}
simulations.  We therefore make no assumptions about the nature of the
convection when the models are mapped into multiple dimensions in {\tt
  MAESTRO}.

%% \begin{figure*}[t]
%% \begin{center}
%% \epsscale{0.9}
%% \plotone{stan_model}
%% \epsscale{1.0}
%% \end{center}
%% \caption{\label{Fig:stan model} {\tt Kepler}-supplemented
%%   \cold\ (solid lines) and \hot\ (dashed lines) models as described in
%%   the text.  Energy release from nuclear burning at the base of the
%%   \He\ layer has caused the temperature to rise.  The \cold\ model is
%%   evolved to a peak $T_\text{base} = 3.67\times10^{8}$ K and the
%%   \hot\ model is evolved to a peak $T_\text{base} = 5.39 \times 10^8$
%%   K.  The black vertical lines indicate the location of the anelastic
%%   cutoff while the grey vertical lines indicate the location of the
%%   beginning of our sponge forcing term for each of the models (see
%%   Section \ref{Sec:Hydrodynamics Algorithm}).}
%% \end{figure*}

%==========================================================================
% Hydrodynamics Algorithm
%==========================================================================
\section{Hydrodynamics Algorithm}\label{Sec:Hydrodynamics Algorithm}
For our multidimensional simulations, we use the low Mach number
stellar hydrodynamics algorithm, {\tt MAESTRO}.  This code is
appropriate for flows in the low Mach number regime, where the
characteristic fluid velocity is small compared to the speed of sound.
Note that the algorithm does not enforce that the Mach number remain
small, but rather is only valid for such flows.  A series of papers
(see \citet{ABRZ:I}---henceforth Paper I, \citet{ABRZ:II}---henceforth
Paper II, \citet{ABNZ:III}---henceforth Paper III, and
\citet{ZABNW:IV}---henceforth Paper IV) describe the derivation of the
low Mach number equation set, its algorithmic implementation, and the
initial application to convection in a white dwarf preceding a Type Ia
supernova.  We use the most recent version of the algorithm, which
includes local adaptive mesh refinement, as given in
\citet{MAESTRO:Multilevel}---henceforth Paper V.

One key advantage of using a low Mach number approach is the increase
of allowable time step size, which enables long-time integration.  Standard
compressible hydrodynamics codes for astrophysical applications such
as {\tt CASTRO} \citep{castro} or {\tt FLASH} \citep{flash} evolve a
fully compressible equation set, i.e., the Euler equations, which
allows for the formation and propagation of shocks.  For low speed
convective motion in our pre-burst convection studies, we do not need
to explicitly follow the propagation of sound waves.  Our low Mach
number equation set does not contain acoustic waves, and therefore
{\tt MAESTRO} is able to take time steps constrained by the maximum
fluid velocity, rather than the maximum sound speed.  As an example,
if the maximum Mach number of the flow is $M \sim 0.01$, we will
obtain a factor of $1/M \sim 100$ increase in time step size compared
to a standard compressible approach.  Another advantage of a low Mach
number method is that the overall HSE of the state can be guaranteed
by the inclusion of a base state in HSE in the low Mach number
equation set.  This removes the difficulties of maintaining HSE
commonly found in compressible hydrodynamics codes.

{\tt MAESTRO} solves a system of advection-reaction-diffusion
equations with the equation of state formulated as an elliptic
constraint on the velocity.  {\tt MAESTRO} uses a higher-order Godunov
method to discretize the advective terms, Strang-splitting to couple
the reaction terms to the advective terms, and a semi-implicit
treatment of the diffusion terms.  The diffusion term and the
divergence constraint are formulated as linear systems which are
solved iteratively using multigrid.  The evolution of the
one-dimensional base state density is also computed.  The base state
density represents the average state of the atmosphere, and is coupled
to the base state pressure via HSE.  The base state density has its
own evolution equation that computes the expansion of the atmosphere
due to heating and is discretized using a higher-order Godunov method.
We note that {\tt MAESTRO} is second-order accurate in space and time.

\subsection{{\tt MAESTRO} Details}\label{Sec:MAESTRO Details}
We now provide additional details of the low Mach number equation set
and numerical implementation.  The interested reader is referred to
Papers I-V for full details.  We use a two-dimensional Cartesian
formulation with $x$ the horizontal coordinate and $r$ the radial
coordinate.  The low Mach number equation set is:
\begin{eqnarray}
\frac{\partial (\rho X_k)}{\partial t} &=& - \nablab\cdot(\rho X_k\Ub)
  + \rho\omegadot_k, \label{eq:Species Equation}\\ 
\frac{\partial\Ub}{\partial t} &=& -\Ub\cdot\nablab\Ub -
  \frac{1}{\rho}\nablab\pi - \frac{(\rho-\rho_0)}{\rho}g\eb_r, 
  \label{eq:Velocity Equation} \\
\frac{\partial(\rho h)}{\partial t} &=& -\nablab\cdot(\rho h \Ub) + 
  \frac{Dp_0}{Dt} + \rho\Hnuc + \nablab\cdot(\kth\nablab T), 
  \label{eq:Enthalpy Equation}
\end{eqnarray}
where $\Ub$, $h$ and $\kth$ are the velocity, specific enthalpy, and
thermal conductivity, respectively.  The species are represented by
their mass fractions, $X_k$, along with their associated production
rates, $\omegadot_k$, and $\Hnuc$ is the nuclear energy generation
rate per unit mass.  Using low Mach number asymptotics (see Paper I)
the total pressure, $p(x,r,t)$, is decomposed into a base state
pressure, $p_0(r,t)$, and a perturbational, or dynamic, pressure,
$\pi(x,r,t)$, such that $|\pi|/p_0 = \mathcal{O}(M^2)$.  The base
state density, $\rho_0(r,t)$, is in HSE with the base state pressure
such that $\nablab p_0 = -\rho_0g\eb_r$, where $\eb_r$ is the unit
vector in the outward radial direction.  

Thermal conduction was not present in Paper V, so we have developed a
semi-implicit discretization for this term.  We include full
algorithmic implementation details in Appendix \ref{Sec:Thermal
  Diffusion} and a verification test problem in Appendix
\ref{Sec:Diffusion Solver Test}.

Mathematically, this system must still be closed by the equation of
state, which is expressed as a divergence constraint on the velocity
field (see Paper III),
\begin{equation}
 \nablab\cdot(\beta_0\Ub) = \beta_0\left(S -
 \frac{1}{\Gammaonebar p_0}\frac{\partial p_0}{\partial
   t}\right)\label{eq:Divergence Constraint in Theory},
\end{equation}
where $\beta_0$ is a density-like variable,
\begin{equation}
\beta_0(r,t) = \rho(0,t)\exp\left({\int_0^r \frac{1}{\Gammaonebar
    p_0}\frac{\partial p_0}{\partial r'}\;dr'}\right),
\end{equation}
and $\Gammaonebar(r)$ is the average of $\Gamma_1 = \left(d\ln p/
d\ln\rho\right)_s$ where the subscript $s$ means the derivative is
taken at constant entropy.  We will use an overline notation to
represent the average of a quantity, which computationally is the
arithmetic average of all grid cells at a particular radius.  The
expansion term, $S$, in (\ref{eq:Divergence Constraint in Theory})
accounts for local compressibility effects resulting from nuclear
burning, compositional changes, and thermal conduction:
\begin{eqnarray}
S &=& \sigma\Hnuc + - \sigma\sum_k\xi_k\omegadot_k + \frac{1}{\rho
  p_\rho}\sum_kp_{X_k}\omegadot_k + 
  \frac{\sigma}{\rho}\nablab\cdot(\kth\nablab T), \label{eq:S Equation}
\end{eqnarray}
where $\xi_k \equiv \left(\partial h/\partial X_k\right)_{\rho,T,(X_j,j\neq
    k)}$, $p_\rho \equiv \left(\partial p/\partial
  \rho\right)_{T,X_k}$, $p_{X_k} \equiv \left(\partial p / \partial
  X_k\right)_{T,\rho,(X_j,j\neq k)}$ and $\sigma \equiv p_T/(\rho c_p
  p_\rho)$ with $p_T \equiv \left(\partial p/\partial T\right)_{\rho,X_k}$, and
  $c_p \equiv \left(\partial h/\partial T\right)_{p,X_k}$.

Another addition to the {\tt MAESTRO} algorithm is the use of a
``volume discrepancy'' correction.  Because (\ref{eq:Divergence
  Constraint in Theory}) is a linearization of the nonlinear
constraint imposed by the equation of state, the thermodynamic
pressure, $p_\text{EOS} = p(\rho,h,X_k)$, may drift from the base
state pressure, $p_0$, \citep{pember_etal:1998}.  To correct for this
drift, (\ref{eq:Divergence Constraint in Theory}) is augmented with a
term that drives the thermodynamic pressure back to the that of the
base state:
\begin{equation}\label{eq:Divergence Constraint in Practice}
  \nablab\cdot(\beta_0\Ub) = \beta_0\left(S-\frac{1}{\Gammaonebar
    p_0}\frac{\partial p_0}{\partial t} - \frac{f}{\Gammaonebar
    p_0}\frac{p_0-p_\text{EOS}}{\dt}\right),
\end{equation}
where $f$ is the volume discrepancy correction factor and $0 \leq f
\leq 1$.  In Section \ref{Sec:Effects of the Volume Discrepancy Term},
we explore the effectiveness of this term at keeping the overall
solution in thermodynamic equilibrium.

To track the evolution of the base state density, we first define the 
expansion velocity as the average outward velocity:
\begin{equation}
w_0(r,t) = \overline{(\Ub\cdot\eb_r)}.
\end{equation}
As described in Paper V, we compute $w_0$ by integrating a one-dimensional
divergence constraint, found by taking the average of equation 
(\ref{eq:Divergence Constraint in Practice}):
\begin{equation}\label{eq:Divergence Constraint on w0}
  \frac{\partial\left(\beta_0w_0\right)}{\partial r} =
  \beta_0\left(\Sbar - \frac{1}{\Gammaonebar p_0}\frac{\partial
    p_0}{\partial t} - \frac{f}{\Gammaonebar p_0}\frac{p_0 -
    \overline{p_\text{EOS}}}{\dt}\right).
\end{equation}
The evolution equation for the base state density can be found by considering
the average of the continuity equation:
\begin{equation}
\frac{\partial\rho_0}{\partial t} = -\nabla\cdot(\rho_0 w_0\eb_r),
\end{equation}
which we discretize with a higher-order Godunov method.

We use special care in dealing with the low density region of our
simulation.  The density spans many orders of magnitude, and due to
conservation of momentum we may generate large velocities in the upper
atmosphere that do not affect the solution in the higher-density
region.  Unfortunately, these large velocities reduce the efficiency
of our method by reducing the time step size.  The first technique we use
to address this problem is the use of a cutoff density, $\rho_{\rm
  cutoff}$, which is the value we hold the density to outside the
star.  The second technique we use is the use of an anelastic cutoff
density, $\rho_{\rm anelastic}$, below which we determine $\beta_0$ by
keeping the ratio $\beta_0/\rho_0$ constant in the divergence
constraint in order to minimize spurious wave generation.  Full
implementation details for the cutoff densities are described in
Appendix A.5 of Paper V.  In this paper, we use $\rho_{\rm cutoff} =
\rho_{\rm anelastic} = 10^4 \text{\ g/cm}^3$.

The third technique adopted for the low density region is sponging (or
damping), which is used to reduce gravity waves at the surface of the
star.  This technique is commonly used in the atmospheric modeling
community as lateral boundary conditions of limited area simulations
(see, for example, \citealt{Kesel_Win72,Perkey_Kreit76}) as well as
upper boundary conditions to reduce wave reflection off of sharp
gradients in the atmospheric structure (see, for example,
\citealt{Durran_Klemp83,Durran_mono,Chen_etal05}).  In addition, we
have previously utilized the sponging technique in the study of
convection in the cores of white dwarfs (Paper IV).  Full details for
the sponge implementation in \texttt{MAESTRO} can be found in Papers
III and IV, but in summary, we add a forcing term to the velocity,
which effectively damps the velocity so that $\Ub^{\text{new}} \to
\Ub^{\text{new}} * f_\text{damp}$.  We use the following formulation
for the sponge:
\begin{equation}\label{eq:sponge}
f_\text{damp} = \left\{\begin{array}{ll} 
    1, & r \le r_{\rm sp},  \\ \noalign{\medskip}
    \frac{1}{2}\left(1-f_{\rm damp,min}\right)
    \cos\left[\pi\left(\frac{r-r_{\rm sp}}{r_{\rm tp}-r_{\rm sp}}\right)\right]
    + \frac{1}{2}\left(1 + f_{\rm damp,min}\right), & r_{\rm sp} < r \le r_{\rm tp}, \\ \noalign{\medskip}
    f_{\rm damp,min}, & r_{\rm tp} < r,
    \end{array}\right.
\end{equation}
where in our simulations $r_{\rm sp}$ is the radius at which $\rho_0 =
25\rho_{\rm cutoff}$, $r_{\rm tp}$ is the radius at which $\rho_0 =
\rho_{\rm cutoff}$, and $f_{\rm damp,min} = 0.01$.\footnote{Note that
  the form of this sponge is similar to that presented in Section
  4.3.1 of Paper III but with $\kappa \dt = 1$ at each time step.}  In
Figure \ref{Fig:stan model}, the vertical grey lines correspond to the
location of $r_{\rm sp}$ and the vertical black lines correspond to
the location of $r_{\rm tp}$ for each of the initial models.  Figure
\ref{Fig:sponge} shows the initial profile of the sponge for the
\cold\ model.  Note that as the system evolves it is free to expand
thus changing the location of the density cutoffs and consequently the
location and extent of the sponge.  The inclusion of a sponge layer
does not strictly conserve kinetic energy in the sponged region.  We
note, however, that the material above the surface of the star is at
relatively low density compared to the material in the convective
region, and therefore the total amount of energy non-conservation is
small.  Furthermore, as shown in Figure 4 of Paper III, the inclusion
of a sponge layer in the low density region of a simulation does not
affect the dynamics of the flow in the convective region of interest.

%% \begin{figure*}[t]
%% \begin{center}
%% \plotone{sponge}
%% \end{center}
%% \caption{\label{Fig:sponge} Sponge profile for the \cold\ model where
%%   $r_{\rm sp} = 680$ cm and $r_{\rm tp} = 844$ cm.}
%% \end{figure*}

%==========================================================================
% Results
%==========================================================================
\section{Results}\label{Sec:Results}
We describe below the results of mapping the {\tt Kepler}-supplemented
models into {\tt MAESTRO} in two dimensions, ($x,\ r$), and the
system's subsequent evolution.  Section \ref{Sec:Resolution
  Requirements} describes the resolution requirements needed to
properly resolve the burning layer.  In Section \ref{Sec:Effects of
  Thermal Diffusion} we show how the inclusion of thermal diffusion
affects the nuclear burning layer and its location.  We show in
Section \ref{Sec:Expansion Due to Heating} how utilizing a
time-dependent base state allows us to capture the expansion of the
atmosphere due to heating.  Section \ref{Sec:Effects of the Volume
  Discrepancy Term} shows how including a volume discrepancy
correction keeps the base state thermally consistent with the equation
of state.  Finally we discuss the extent and evolution of the
convective region in Section \ref{Sec:Convection}.

To map the one-dimensional model into {\tt MAESTRO}, we 
copy laterally across the domain such that $\phi(x,r,t=0) =
\phi_{\text{one-d}}(r)$ for each variable $\phi$ in the model.
In the following analysis, the subscript ``max'' refers to the maximum
value of a quantity in the computational domain at a given time step.
In two dimensions, we define the average as a function of radius,$\avgtwod{\phi} = \avgtwod{\phi}(r,t)$, of a quantity $\phi$ by
\begin{equation}\label{eq:lateral average 2d}
  \avgtwod{\phi}_j^m = \frac{1}{N}\displaystyle\sum_{i=1}^{N}\phi_{i,j}^m
\end{equation}
where $\phi_{i,j}^m = \phi(x_i,r_j,t^m)$ and $N$ is the total number
of grid zones in the lateral, $x$, direction at height $r_j$ at time
$t^m$.

We use the general equation of state of \cite{timmes_swesty:2000},
which includes contributions from electrons, ions, and radiation.  We
calculate opacities using Frank Timmes' publicly available
conductivity routine, which includes contributions from radiation and
electron conduction processes as explained in \cite{Timmes00}.  It is
important to note that the method for calculating opacities used in
{\tt MAESTRO} is not the same as what is used the {\tt Kepler} code;
it is likely, however, that the different methods give opacities that
agree to within a factor of $\sim 2$ \citep[see discussion
  in][]{HEGER_ETAL07B}.  The boundary conditions for all simulations
are periodic in the $x$-direction to mimic a laterally extended
convection region.  The upper $r$ boundary is outflow to allow for
free expansion of the atmosphere.  The lower $r$ boundary of dense
neutron star material is set to a wall with no normal flow.  To solve
the thermal diffusion contribution at the upper and lower boundaries
we impose the Neumann condition $dh/d\boldsymbol{n}=0$, where
$\boldsymbol{n}$ is the outward facing normal vector; the enthalpy
boundary conditions are periodic in the lateral directions.  We note
that the upper and lower domain boundaries are sufficiently far from
the burning layer so that they do not affect the dynamics of the
convection.  An advective CFL number of $0.7$ was used in all of our
simulations.

As previously mentioned, we do not obtain any multidimensional
velocity information from the {\tt Kepler} models; our system is
initially static.  For convection to begin, the symmetry of the system
must, therefore, be broken.  This can be accomplished either by
placing a small perturbation at the base of the \He\ layer or by
allowing numerical noise from the multigrid solver to seed the
convective cells.  For the simulations presented here, neither
approach is advantageous over the other, both giving quantitatively
similar steady-state convective flow fields; we utilize both
approaches in our studies and when perturbing we place a small
($\Delta T/T = 10^{-5}$) Gaussian temperature perturbation laterally
centered at height $r = 384$ cm to break the initial symmetry of the
problem.

\subsection{Resolution Requirements}\label{Sec:Resolution Requirements}
To date, the only other paper in the literature regarding
multidimensional simulations of XRBs as deflagrations \citep{Lin:2006}
used a finest resolution of $5$ cm zone$^{-1}$.  They presented
multidimensional results at $5$, $7.5$, and $10$ cm zone$^{-1}$
resolutions and remarked that there is a ``tendency toward convergence
with increasing resolution'' based on the time to reach the peak
energy generation rate.  It is important to note that our initial
models are different from those of \cite{Lin:2006}.  In particular,
their models only considered two species---the accreted layer was pure
\He\ and the underlying neutron star was composed entirely of \C.
This caused their models to have a smaller jump in mean molecular
weight across the neutron star/accreted layer boundary compared to our
models.  Furthermore, the initial conditions for their
multidimensional studies were from the results of a one-dimensional
diffusional-thermal code that evolved the system through several
bursts.  These differences from our method of initial model generation
give the \cite{Lin:2006} models an extended ($\sim 100$ cm) thermal
peak compared to our narrow ($\sim 10$ cm) peak (compare our Figure
\ref{Fig:stan model} to their Figure 2).

The burning layer at the base of the accreted material is very thin;
high resolution is required to properly model this region.  The peak
of the thermal profile for the \hot\ model is broader than the
corresponding peak in the \cold\ model.  Consequently, the burning
layer in the \hot\ model is thicker than that of the \cold\ model---we
therefore focus our study of resolution requirements on the more
restrictive of the two, the \cold\ model.  The top panel of Figure
\ref{Fig:enuc at 1ms} shows the $\avgtwod{\Hnuc}$ profile at $t = 1$
ms for simulations of the \cold\ model using the same resolutions as
in the \cite{Lin:2006} study.  Even at this early time there is a
$25$\% spread in the peak value of $\avgtwod{\Hnuc}$ for these
resolutions.  The bottom panel shows the same profile but at several
higher resolutions.  The peak value of $\avgtwod{\Hnuc}$ for the $4$
cm zone$^{-1}$ simulation is comparable to the peak values in the top
panel, but we only see numerical convergence of the peak value as we
go to higher spatial resolution.  In addition, the shape of the
profile near peak converges with increasing resolution; the $0.25$ and
$0.5$ cm zone$^{-1}$ resolution simulations look qualitatively
similar.  We therefore claim that the burning layer is not properly
resolved in our models unless a resolution of $0.5$ cm zone$^{-1}$ is
used.  It is important to note that even though our initial models
differ, this resolution requirement is an order of magnitude higher
than what has been previously presented in the literature and
therefore significantly increases the computational cost of our XRB
simulations.

%% \begin{figure*}[th]
%% \begin{center}
%% \epsscale{0.9}
%% \plotone{enuc_1ms}
%% \epsscale{1.0}
%% \end{center}
%% \caption{\label{Fig:enuc at 1ms} Average of $H_\text{nuc}$ as
%%   a function of height for the \cold\ model at various resolution
%%   models at $t = 1$ ms.  Note that the vertical axes of the inset plots
%%   are in a logarithmic scale.  For clarity, the top panel shows the
%%   same resolutions used in Figure \ref{Fig:coarse evolution of enuc}
%%   and the bottom panel shows more resolved simulations.  The peak of
%%   the profile at $0.5$ cm resolution is qualitatively similar to the
%%   peak of the profile at $0.25$ cm resolution. }
%% \end{figure*}

Under-resolving the burning layer artificially boosts the energy
generation rate, which in turn over-drives convection.  Figure
\ref{Fig:under-resolved convection} shows a close-up of the \C\ mass
fraction after $10$ ms of evolution of the \cold\ model at $0.5$ ({\tt
  a}), $2$ ({\tt b}), $4$ ({\tt c}), and $7.5$ cm zone$^{-1}$ ({\tt
  d}) resolutions.  The base of the burning layer is located in the
bottom-most green region (just below the magenta) in panel {\tt a}.
All four simulations give a well-mixed carbon region above the burning
layer; the extent of the convective zone increases with decreasing
resolution with the $7.5$ cm zone$^{-1}$ simulation's convective zone
extending $30\%$ further than the $0.5$ cm zone$^{-1}$ simulation's
convective zone.  The amount of convective undershoot---the tendency
of material to penetrate below the burning layer---is much more
sensitive to resolution.  The $0.5$ cm zone$^{-1}$ simulation shows
very little evidence of undershooting while the $7.5$ cm zone$^{-1}$
simulation has an undershoot region that is larger in physical extent
than its corresponding convective region above the burning layer.  For
all of the studies described below, we use a resolution of $0.5$ cm
zone$^{-1}$ in the burning layer.

%% \begin{figure*}[th]
%% \begin{center}
%% \epsscale{0.9}
%% \plotone{c12_convection2}
%% \epsscale{1.0}
%% \end{center}
%% \caption{\label{Fig:under-resolved convection} Effects of
%%   under-resolving convection for the \cold\ model. Plotted is the
%%   \C\ mass fraction after $10$ ms of evolution for various
%%   resolutions: {\tt a}) $0.5$, {\tt b}) $2$, {\tt c}) $4$ and {\tt d})
%%   $7.5$ cm zone$^{-1}$.  Each figure shows the same region of physical
%%   space and has dimensions $256$ cm $\times$ $1024$ cm.  The coarse
%%   resolution simulations show an extended convective zone and a
%%   significant amount of convective undershoot. }
%% \end{figure*}

\subsection{Effects of Thermal Diffusion on the Burning Layer}\label{Sec:Effects of Thermal Diffusion}
As explained in Section \ref{Sec:Introduction}, the burning front
during an XRB likely propagates as a subsonic flame, the speed of
which is regulated by the rate of thermal diffusion across the front.
At the resolution required to resolve the thin burning layer (see
previous section) it is currently intractable to evolve the system
until flame ignition.  We can, however, investigate the effects of
thermal diffusion on the stable burning that occurs in the burning
layer.  Here we focus on the \hot\ model instead of the \cold\ model
because it has the larger thermal gradient---and hence diffusive heat
flux---at the base of the accreted layer.  For this simulation, we use
the new adaptive mesh refinement (AMR) capability in {\tt MAESTRO}
\citep{MAESTRO:Multilevel}, using 2 levels of refinement and ensuring
that the entire convective region is at the finest level of refinement
with resolution $0.5$ cm zone$^{-1}$.  Figure \ref{Fig:effect of
  diffusion} shows these effects in $\mymax{\Hnuc}$ (solid lines) and
the location of this maximum (dashed lines) as a function of time at
early times both with (green) and without (blue) thermal diffusion.
We note that the location of $\mymax{\Hnuc}$ is always at the finest
level of refinement.  The $\mymax{\Hnuc}$ evolution is similar for
both cases with the magnitude in general being slightly larger for the
case of no diffusion.  The initial spike in $\mymax{\Hnuc}$ at $t
\approx 0.25$ ms is due to the fact that, initially, there is no
established fluid flow that can advect away the energy released from
nuclear reactions (see the discussion in Section
\ref{Sec:Convection}).  Over the next $3$ ms, the location of
$\mymax{\Hnuc}$ for both simulations moves radially inward at a rate
of $\sim2.9\times10^3$ cm s$^{-1}$.  Around $t=3.25$ ms, the inward
radial progression of the location of $\mymax{\Hnuc}$ for the
simulation with no diffusion significantly slows to $\sim 900$ cm
s$^{-1}$.  For the remainder of the simulation, the case with thermal
diffusion shows no such slowdown---heat transported radially inward
via diffusion expands the lower boundary of the convective zone, which
mixes fresh fuel to slightly deeper layers.  By the end of the
simulations, the case that included diffusion had an $\mymax{\Hnuc}$
that occurred $\sim 4$ cm deeper within the atmosphere than in the
case without diffusion.  It should be noted that the typical standard
deviation in the location of $\mymax{\Hnuc}$ for the case without
diffusion is of order $2$ cm; this suggests that perhaps thermal
diffusion plays a role in regulating the location of maximum nuclear
burning, but further evolution is needed to make statistically
significant claims.

%% \begin{figure*}[th]
%% \begin{center}
%% \plotone{diff_vs_nodiff}
%% \end{center}
%% \caption{\label{Fig:effect of diffusion}Evolution of $\mymax{\Hnuc}$
%%   (solid lines) and its vertical location (dashed lines) as a function
%%   of time for the \hot\ model both with (green) and without (blue)
%%   thermal diffusion.  }
%% \end{figure*}

\subsection{Expansion of Base State due to Heating}\label{Sec:Expansion Due to Heating}
Having a dynamical base state allows us to capture the large-scale
expansion of the atmosphere due to heating from nuclear reactions.
This differs from the work by \cite{Lin:2006}, which had a
time-independent base state and did not model the top of the accreted
atmosphere due to numerical complications with their algorithm.
Figure \ref{Fig:expansion of base state} shows the ratio of the base
state density to that of the initial ($t=0$) base state density
profile near the surface of the atmosphere for the \cold\ model.  We
define the surface to be where $\rho_0 = \rho_{\rm cutoff}$.  The
vertical dashed lines represent the location of the surface for each
time-value.  After $26.6$ ms of evolution, the base state has
responded to heating from nuclear reactions approximately $4.5$ m
below the surface by expanding $3.5$ cm.  The lower Mach number flow
in the \cold\ model compared to the \hot\ model allows for longer-term
evolution of the system and therefore larger expansion of the
atmosphere.

The extent of the expansion is rather small at these early times.
However, as the system progresses towards outburst the energy
generation and, therefore, the rate of expansion increases.  As the
system expands, the $p_0$ profile changes, which can affect the
dynamics in the convective region.  Additionally, as the atmosphere
expands, the burning layer becomes less degenerate, which may be
important for the nucleosynthesis during the outburst.  Furthermore, a
proper modeling of this expansion during the peak of a PRE burst model
may help pinpoint the location of the photosphere with respect to the
stellar radius at touchdown, a quantity that plays an important role
in using XRBs to measure the mass and radius of the underlying neutron
star \citep[for example]{STEINER_ETAL10}.

%% \begin{figure*}[th]
%% \begin{center}
%% \plotone{base_state_comparison}
%% \end{center}
%% \caption{\label{Fig:expansion of base state} Expansion of the base
%%   state due to heating.  Plotted is the ratio of base state density to
%%   the initial ($t=0$) base state density near the surface of the
%%   atmosphere for the \cold\ model.  We define the surface to be where
%%   $\rho = \rho_{\rm cutoff}$ and it is represented by the vertical lines.
%%   The base state has expanded $3.5$ cm in $26.6$ ms of evolution. }
%% \end{figure*}

\subsection{Effects of the Volume Discrepancy Term}\label{Sec:Effects of the Volume Discrepancy Term}
In Section \ref{Sec:MAESTRO Details} we explained that the
thermodynamic pressure may drift from the base state pressure.  To
correct for this drift, we introduced the volume discrepancy term in
equation (\ref{eq:Divergence Constraint in Practice}), which drives
the thermodynamic pressure towards the base state pressure.  We focus
our attention here on the \hot\ model because it shows a more dramatic
drift of the thermodynamic pressure from the base state pressure.
Figure \ref{Fig:volume discrepancy effect} shows the volume
discrepancy term in action by examining the percent difference between
the base state and thermodynamic pressures as a function of time for
various values of $f$ for the \hot\ model.  The top panel shows the
maximum value whereas the bottom panel shows the average value of this
percent difference; both the peak and average values show the same
trend for a given value of $f$.  After the initial adjustment of the
system, the average drift for the case of no volume discrepancy
correction ($f=0$) increases approximately linearly at $\sim 0.1\%$
per ms of evolution.  Including the correction term restricts the
temporal- and spatial-averaged value of the drift to $\lesssim 0.02
\%$.

For nonzero $f$, the oscillatory behavior in the drift is due to the
fact that the system may slightly over-correct the thermodynamic
pressure in a given time step and then recover in the next step.  A
larger value of $f$ causes a stronger driving of the drift, which
tends to be more oscillatory.  In addition, a larger value of $f$
appears to be correlated with larger spikes in the drift.  The top
panel of Figure \ref{Fig:deltap spike closeup} shows a closeup of the
$\mathcal{O}(1)$ error seen in the $f=0.3$ curve in Figure
\ref{Fig:volume discrepancy effect}.  The location of the maximum
drift is also plotted in the top panel; the large spike in the drift
occurs just below the burning layer at $r = 366.25$ cm.  The bottom
panel of Figure \ref{Fig:deltap spike closeup} shows the corresponding
maximum energy generation rate, which also contains a spike that is
coincident with the spike in the drift---a large deposit of energy on
a short timescale causes the thermodynamic pressure to get out of sync
with the hydrostatic base state pressure.  The increase in the energy
generation rate is due to a fluid parcel rich in \He\ fuel being
brought into a region of high temperature via the turbulent
convection.  The duration of this transient behaviour is very short:
$9$ time steps or $\sim 6.4 \times 10^{-7}$ s.  The selection of an
appropriate non-zero value for $f$ is a problem-specific endeavor, but
the chosen value has little effect on the dynamics of the convective
flow field.  For the simulations presented below we use a volume
discrepancy correction value of $f = 0.3$, which is based on the
results of several test runs and past experience with comparing the
results to the $f=0$ case.  We will continue to study if and how the
chosen value of $f$ affects the long term development of the
convective field for this specific problem.

%% \begin{figure*}[th]
%% \begin{center}
%% \plotone{deltap}
%% \end{center}
%% \caption{\label{Fig:volume discrepancy effect} Effects of volume
%%   discrepancy factor as characterized by the percent difference
%%   between the pressure as given by our equation of state,
%%   $p_\text{EOS}$, and the base state pressure, $p_0$, for the
%%   \hot\ model.  The left panel shows the maximum value whereas the
%%   right panel shows the average value of the percent difference in the
%%   computational domain.  Note the different vertical scale between the
%%   two plots.}
%% \end{figure*}

%% \begin{figure*}[th]
%% \begin{center}
%% \epsscale{0.9}
%% \plotone{deltap_spike_closeup}
%% \epsscale{1.0}
%% \end{center}
%% \caption{\label{Fig:deltap spike closeup} Closeup of the
%%   $\mathcal{O}(1)$ spike in the maximum value of the $f=0.3$ drift as
%%   seen in the left panel of Figure \ref{Fig:volume discrepancy
%%     effect}.  The top panel shows the drift value and its location in
%%   the domain; the bottom panel shows the maximum energy generation
%%   rate.  The large amount of energy released from the burning spike
%%   causes the thermodynamic pressure to differ from the hydrostatic
%%   base state pressure and therefore a spike in the drift.}
%% \end{figure*}

\subsection{Convective Dynamics}\label{Sec:Convection}
The adiabatic excess, $\Delta\nabla$,---with
\begin{equation}\label{eq:adiabatic excess}
  \Delta\nabla = \nabla - \nabla_s,
\end{equation} 
where the actual thermal gradient is
\begin{displaymath}
  \nabla = \frac{\partial \ln T / \partial r}{\partial \ln p / \partial r}
\end{displaymath}
and the adiabatic thermal gradient is $\nabla_s = \left(d\ln T/ d\ln
p\right)_s$ with the subscript $s$ meaning along an adiabat---is used
to gauge the evolution of the convective zone for the Schwarzschild
instability criterion.  Under this criterion, a fluid element is
unstable to thermally driven convection when $\Delta\nabla > 0$ and is
stable for $\Delta\nabla < 0$.  The first term in (\ref{eq:adiabatic
  excess}) is calculated using finite differences of the temperature
and pressure profiles along the radial direction.  The second term in
(\ref{eq:adiabatic excess}) depends solely on the thermodynamics of
the equation of state.  It is related to the second adiabatic
exponent, $\Gamma_2$ (see \cite{CoxGiuli} Chapter 9):
\begin{equation}\label{eq:gamma2}
  \frac{\Gamma_2-1}{\Gamma_2} = \left(\frac{d\ln T}{d\ln p}\right)_s.
\end{equation}
All three of the adiabatic exponents are related:
\begin{equation}\label{eq:adiabatic exponents}
\frac{\Gamma_1}{\Gamma_3 - 1} = \frac{\Gamma_2}{\Gamma_2-1},
\end{equation}
where $\Gamma_3-1 = \left(d\ln T/d\ln \rho\right)_\text{s}$ and
$\Gamma_1$ was defined in Section \ref{Sec:MAESTRO Details}.  Writing
the equation of state as $p = p(\rho,T)$ and expanding the
differential $dp$, we find the relation
\begin{equation}\label{eq:gamma3-gamma1 relation}
\Gamma_3 - 1 = \frac{\Gamma_1 - \rho p_\rho / p}{Tp_T/p}
\end{equation}
along an adiabat.  Our equation of state only returns $\Gamma_1$, but
combining this with (\ref{eq:adiabatic exponents}) and
(\ref{eq:gamma3-gamma1 relation}) allows us to solve for the adiabatic
thermal gradient and hence the adiabatic excess.

Figure \ref{Fig:early adiabatic excess evolution} shows the early
evolution of $\Delta\nabla$ for the \cold\ model.  Each plot covers
the spatial range $(0 \leq x \leq 256$ cm$, 350$ cm $\leq r \leq 700$
cm$)$ to focus on the convective region.  The stripes in the initial
conditions, Figure \ref{SubFig:ad excess 0}, are due to small
interpolation errors from mapping the initial data onto the
two-dimensional grid.  The initial adjustment of the system seen in
Figure \ref{SubFig:ad excess 0.4} causes a mixing of stable (blue) and
unstable (red) fluid elements. This transient adjustment phase occurs
for two reasons: 1) the initial conditions were based on a
parameterization of convection in one dimension and the system now
needs to adjust to a two-dimensional convective zone, and 2) the
initial perturbation does not have an established convection zone and
the system needs a short amount of time to build up a flow pattern
associated with the perturbation.  This results in mixing that
produces a region that is marginally convective ($\Delta\nabla\sim0$;
white) with localized pockets of stable and unstable fluid elements as
seen in Figure \ref{SubFig:ad excess 0.8}.  At later times, these
pockets further localize into vortices whose circulation gives rise to
roughly circular regions of nonzero adiabatic excess---with one
hemisphere that is stable and the other which is unstable---that are
advected with the flow before dispersing into the ambient medium on
subconvective timescales, $\sim10^{-4}$ s.  The vortices are always
associated with an adiabatic excess pattern that has an unstable (red)
bottom and a stable (blue) top unless two vortices are merging and
interacting, in which case the stability distribution becomes skewed.

Figure \ref{Fig:late adiabatic excess evolution} shows $\Delta\nabla$
for the same simulation as in Figure \ref{Fig:early adiabatic excess
  evolution} but at later times.  The boxes in these plots outline a
single long-lived vortex that forms around $t=18.5$ ms, Figure
\ref{SubFig:ad excess 18.5}, and lasts throughout the remainder of the
simulation.  Formation of this vortex is correlated with the formation
of stronger filamentary structures, which are most clear in Figures
\ref{SubFig:ad excess 25}, \ref{SubFig:ad excess 26} and
\ref{SubFig:ad excess 28}.  These filaments appear to wrap around the
solitary vortex and restrict the main formation of smaller vortices to
the lower boundary of the convective region.

\begin{figure*}[t]
\begin{center}
\epsscale{0.9}
\plotone{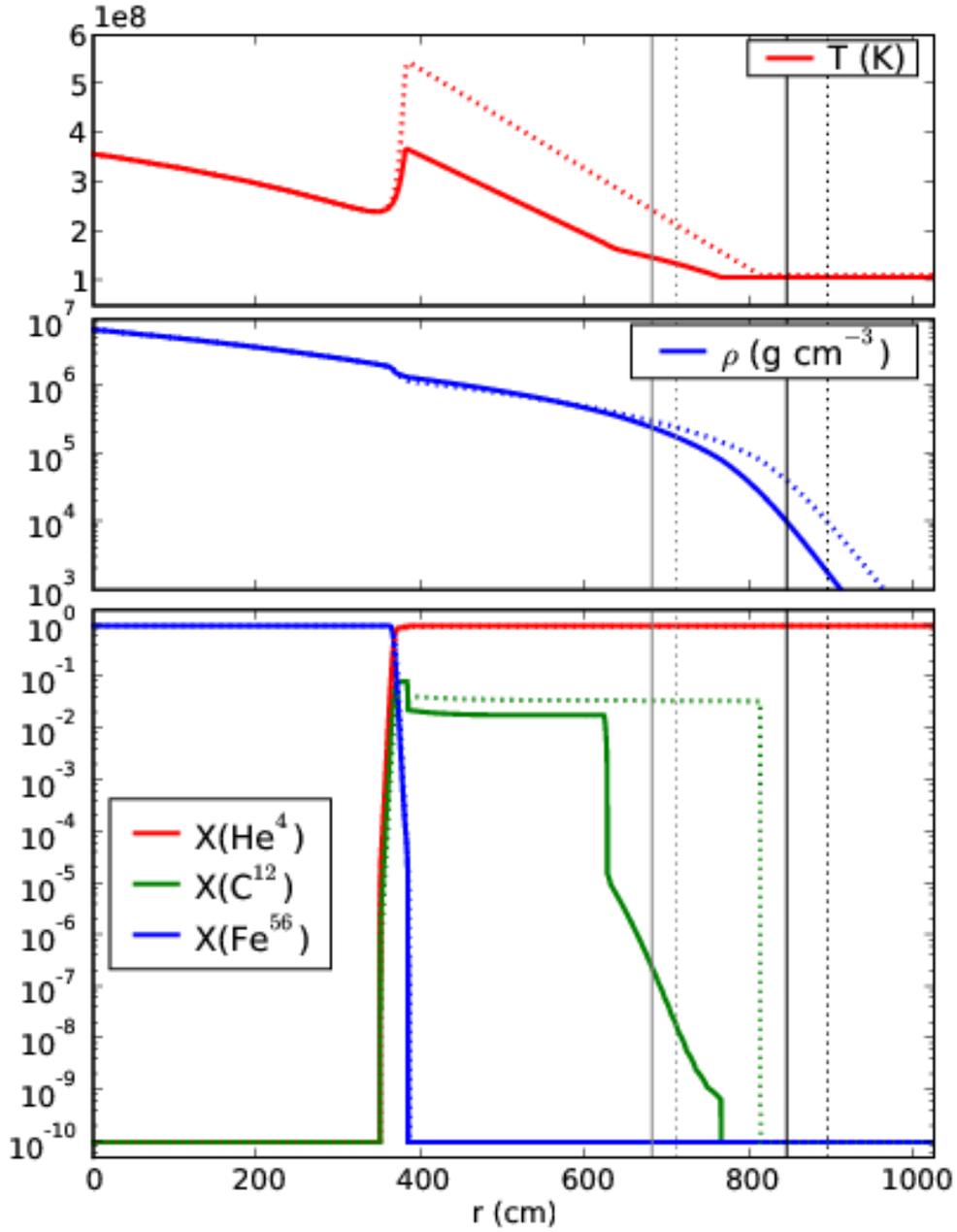}
\epsscale{1.0}
\end{center}
\caption{\label{Fig:stan model} {\tt Kepler}-supplemented
  \cold\ (solid lines) and \hot\ (dashed lines) models as described in
  the text.  Energy release from nuclear burning at the base of the
  \He\ layer has caused the temperature to rise.  The \cold\ model is
  evolved to a peak $T_\text{base} = 3.67\times10^{8}$ K and the
  \hot\ model is evolved to a peak $T_\text{base} = 5.39 \times 10^8$
  K.  The black vertical lines indicate the location of the anelastic
  cutoff while the grey vertical lines indicate the location of the
  beginning of our sponge forcing term for each of the models (see
  Section \ref{Sec:MAESTRO Details}.)}
\end{figure*}

\begin{figure*}[t]
\begin{center}
\plotone{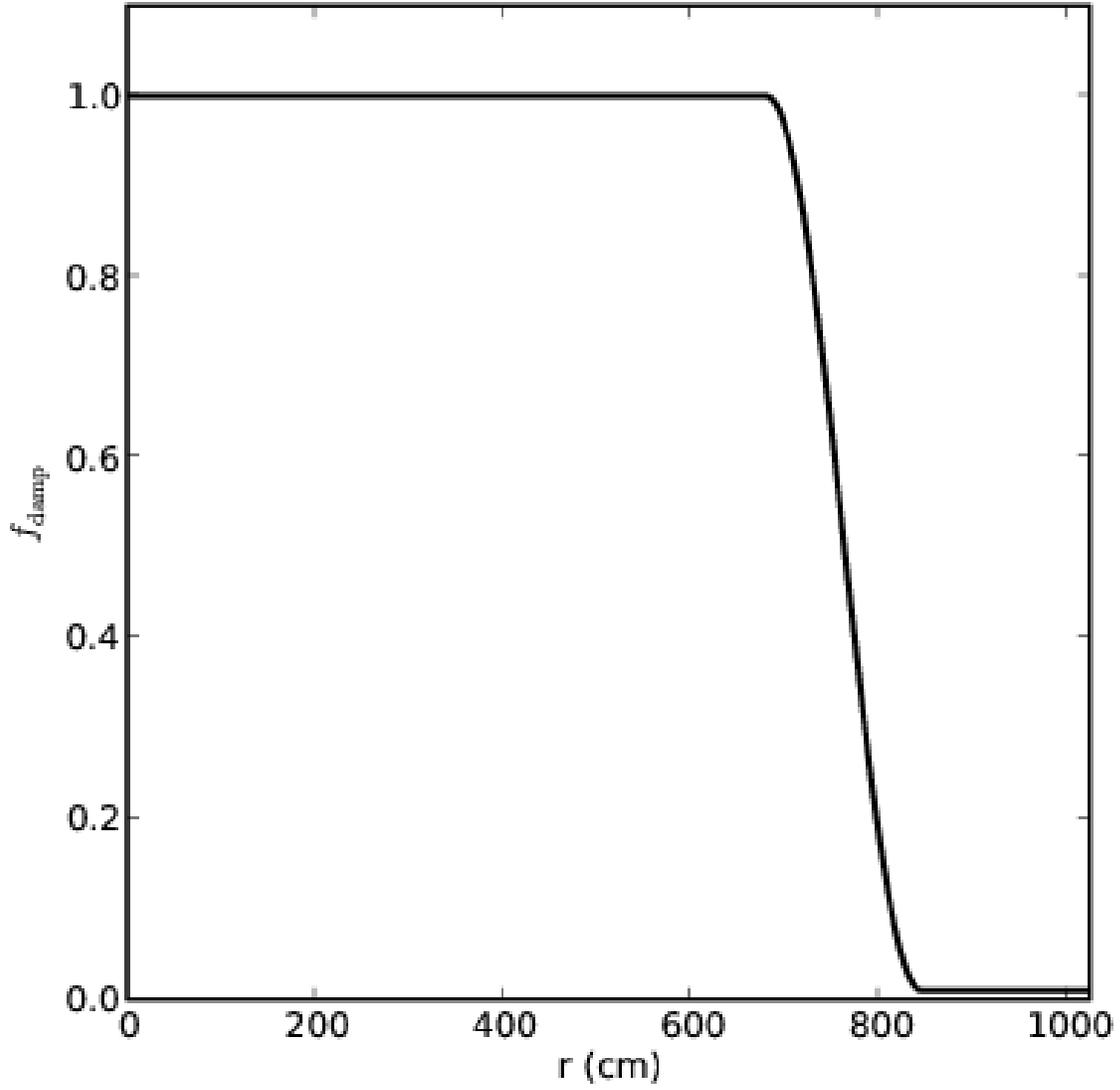}
\end{center}
\caption{\label{Fig:sponge} Initial sponge profile for the \cold\ model where
  $r_{\rm sp} = 680$ cm and $r_{\rm tp} = 844$ cm.}
\end{figure*}

\begin{figure*}[th]
\begin{center}
\epsscale{0.9}
\plotone{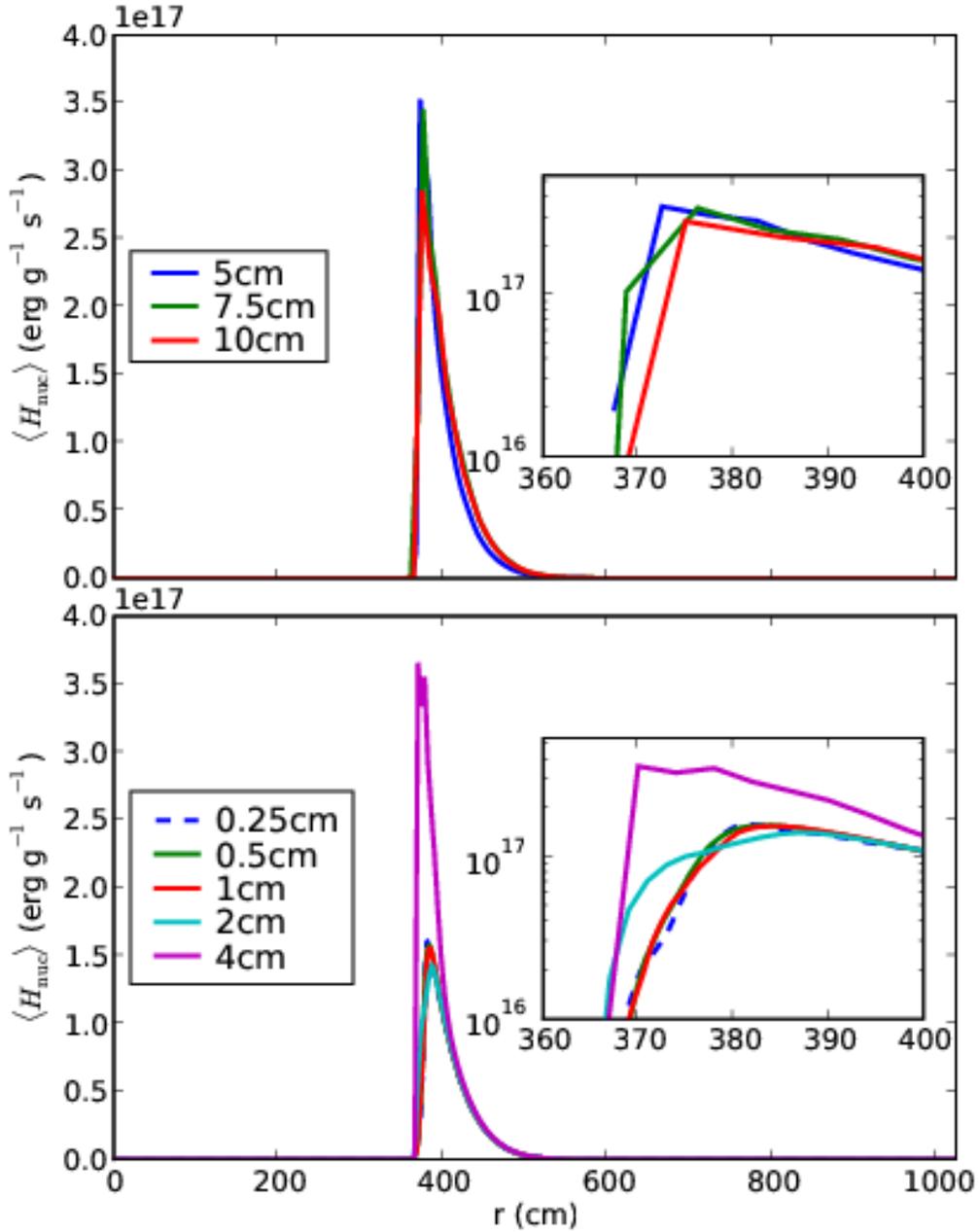}
\epsscale{1.0}
\end{center}
\caption{\label{Fig:enuc at 1ms} Average of $H_\text{nuc}$ as a
  function of height for the \cold\ model at various resolutions at $t
  = 1$ ms.  Note that the vertical axes of the inset plots are in a
  logarithmic scale.  For clarity, the top panel shows simulations
  which use the same resolutions as in the \cite{Lin:2006} study and
  the bottom panel shows more resolved simulations.  The peak of the
  profile at $0.5$ cm zone$^{-1}$ resolution is qualitatively similar
  to the peak of the profile at $0.25$ cm zone$^{-1}$ resolution. }
\end{figure*}

\begin{figure*}[th]
\begin{center}
\epsscale{0.9}
\plotone{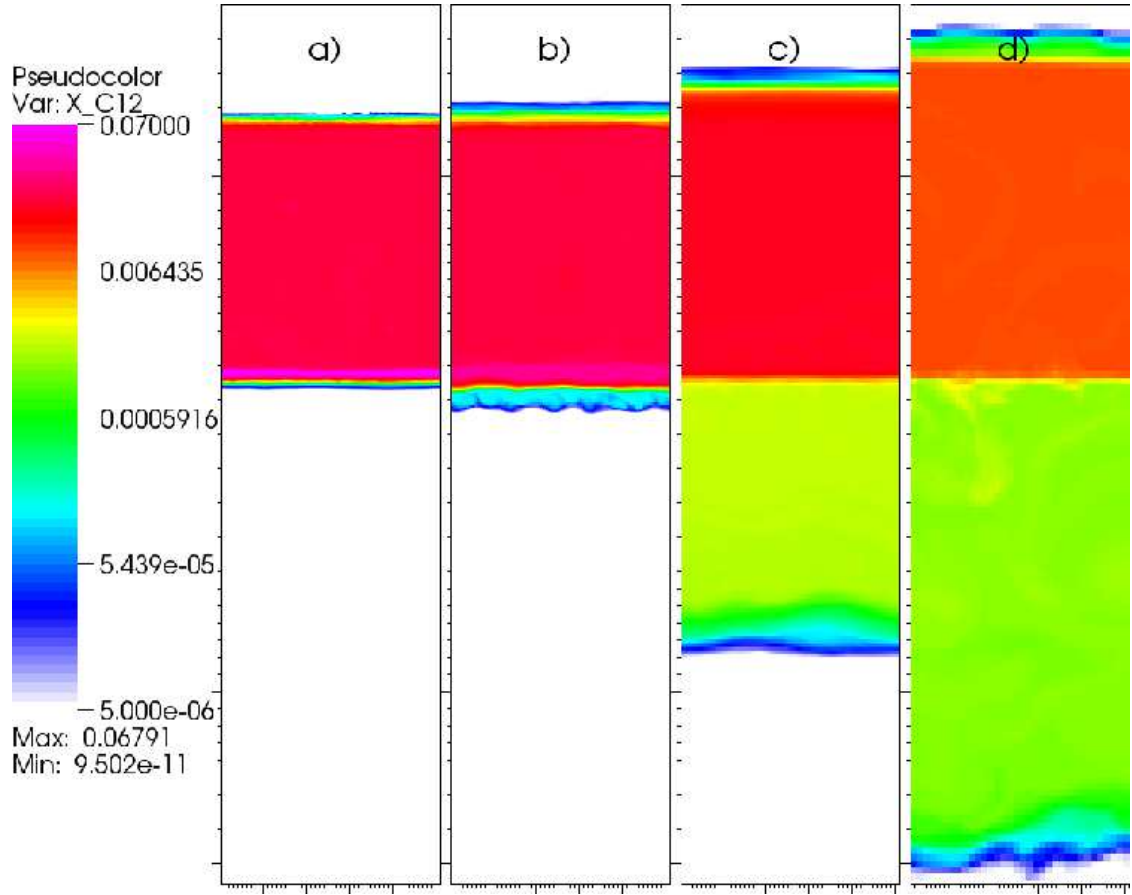}
\epsscale{1.0}
\end{center}
\caption{\label{Fig:under-resolved convection} Effects of
  under-resolving convection for the \cold\ model. Plotted is the
  \C\ mass fraction after $10$ ms of evolution for various
  resolutions: {\tt a}) $0.5$, {\tt b}) $2$, {\tt c}) $4$, and {\tt d})
  $7.5$ cm zone$^{-1}$.  Each figure shows the same region of physical
  space and has dimensions $256$ cm $\times$ $1024$ cm.  The coarse
  resolution simulations show an extended convective zone and a
  significant amount of convective undershoot. }
\end{figure*}

Another way to quantify the convective region is to look at the ratio
$\nabla / \nabla_s$.  From (\ref{eq:adiabatic excess}) we see that the
system is unstable to convection under the Schwarzschild criterion
when $\nabla / \nabla_s > 1$.  The Schwarzschild criterion, however,
does not consider the effects of composition gradients that may help
stabilize the material against convection; for this we need to
consider the Ledoux criterion for instability 
\begin{equation}\label{eq:ledoux criterion}
  \nabla - \nabla_L > 0,
\end{equation}
where the Ledoux thermal gradient is (see, for example,
\cite{Kipp_Weigert})
\begin{displaymath}
  \nabla_L = \nabla_s - \sum_i\frac{\partial \ln X_i / \partial
    r}{\partial \ln p /\partial r}
\end{displaymath}
and the second term above is evaluated via finite differences of the
composition and pressure profiles.  As with the Schwarzschild
criterion, one can look at the ratio $\nabla / \nabla_L$, which is
greater than unity if the material is unstable to Ledoux convection.
Figure \ref{SubFig:example convective profiles} shows the above ratios
for the average thermal gradients for the initial configuration (left)
and after the system has evolved for $t = 23.5$ ms (right); the black
line is for the case of Schwarzschild criterion convection, while the
red line is for Ledoux convection.  The dashed horizontal line marks
the boundary for stability against convection.  Where the curves lie
above this line, the configuration is unstable; when convection is
efficient, the curves should lie very near the horizontal line.  For
both the initial condition and at late times, the Schwarzschild curve
and the Ledoux curve are well matched except near the edges of the
convective region where composition gradients cause the two curves to
deviate slightly.  This is most noticeable in the initial
configuration at the upper boundary where there is a sharp jump in
composition, which was not smoothed (see Figure \ref{Fig:stan model}).
Of interest in the plot at $t = 23.5$ ms is the feature at $r = 450$
cm, which has an unstable bottom and a stable top; this is consistent
with the vortices in Figures \ref{Fig:early adiabatic excess
  evolution} and \ref{Fig:late adiabatic excess evolution}, which had
red bottoms and blue tops.  We define the edge of the convective
region to be where $\nabla / \nabla_s$, $\nabla / \nabla_L = 0.75$.
This particular value of $0.75$ was chosen to be sufficiently small
enough to rule out false positives from strong pockets of stability
from, for example, vortices within the convective region, but also
large enough to rule out any fluctuations at the boundaries due to
overshoot.  Figure \ref{SubFig:extents of convection} shows in grey
the extent of the convective region as a function of time with respect
to the full domain for both the Schwarzschild (left) and Ledoux
(right) instability criteria.  The horizontal dashed lines mark the
initial location of the lower and upper boundaries.  The overall
expansion of the upper boundary for the Schwarzschild (Ledoux)
criterion is $32.0$ ($29.5$) cm in $30$ ms of evolution; the lower
boundary expands downward by $9.5$ ($6.0$) cm in the same time.  At
late times, the upper boundary of the convective region has a much
smoother composition transition than the lower boundary, therefore,
the Schwarzschild and Ledoux criteria are much better matched at the
upper boundary than the lower.  Nevertheless, both the Ledoux and
Schwarzschild criteria yield similar results when used to determine
the extent of the convective region.  In terms of column-depth, the
convective zone after $30$ ms of evolution spans the region $2.2\times
10^7$ g cm$^{-2} \lesssim y \lesssim 2.6\times10^8$ g cm$^{-2}$ for
both instability criteria.

%% \begin{figure}[th]
%%   \centering
%%   \subfigure[Example Convective Profiles]{
%%     \plotone{examples}
%%     \label{SubFig:example convective profiles}
%%   }
%%   \subfigure[Convective Region Extent]{
%%     \plotone{extents}
%%     \label{SubFig:extents of convection}
%%   }
%%   \caption{\label{Fig:convective boundaries}Analysis of the extent of
%%     the convective region.  Panel (a) shows the convective profiles
%%     for both the Schwarzschild and Ledoux instability criteria at two
%%     different times.  Panel (b) shows the extent of the convective
%%     region as a function of time as determined by both instability
%%     criteria.
%% \end{figure}

For comparison, Figures \ref{Fig:c12 early} and \ref{Fig:c12 late}
show the \C\ mass fraction with velocity vectors for the same
simulation and at the same times as in Figures \ref{Fig:early
  adiabatic excess evolution} and \ref{Fig:late adiabatic excess
  evolution}, respectively.  These figures clearly show the
association of vortices with the circular regions of nonzero adiabatic
excess seen in Figures \ref{Fig:early adiabatic excess evolution} and
\ref{Fig:late adiabatic excess evolution}.  The initial adjustment of
the system causes mixing that smooths the slight over-abundance of
\C\ at the base of the accreted layer present in the initial model
(see Figure \ref{Fig:stan model}).  At late times, the convective
region is very well-mixed, and the \C\ mass fraction is nearly
laterally homogeneous.  Furthermore, the circulation pattern
associated with the long-lived vortex outlined in Figure \ref{Fig:late
  adiabatic excess evolution} has grown to a large fraction of the
convective zone and is self-interacting because of the periodic
boundary conditions.  The tendency of the system to form a single
dominant vortex from smaller vortices is a feature of two-dimensional
simulations.  In three dimensions, the turbulent energy cascade moves
from large to small scales; large vortical structures break down into
smaller structures that are eventually dissipated by viscous effects.
In two dimensions, as is the case here, the turbulent energy cascade
is reversed---small vortical structures merge together to form a
single dominant vortex.  

The circulation is counter-clockwise for the large, long-lived vortex;
this causes a region with positive $x$-velocity below and a region of
negative $x$-velocity above the vortex center.  The positive
$x$-velocity region extends all the way to the lower convective
boundary where it causes shearing of the \He/\C-rich region with the
underlying \Fe\ region.  Consequently, Figure \ref{Fig:iron
  enrichment} shows that some of the underlying \Fe\ neutron star
material is churned up into the convective region where it is mixed
with the rest of the convective material.  The left panel shows
average \Fe\ mass fraction profiles starting with the initial model
abundance (thick solid line) through $t = 30$ ms (thick dashed line);
the intermediate thin solid lines show profiles at the times used in
Figures \ref{Fig:early adiabatic excess evolution} and \ref{Fig:late
  adiabatic excess evolution}.  By $t = 5$ ms, the \Fe\ is fairly
well-mixed in the convective region.  The right panel shows the total
mass of \Fe\ in the region defined by the initial convective zone.
The greatest growth in the total mass occurs, as expected, in the
initial adjustment ($t \lesssim 0.6$ ms) and then flattens until large
enough structures form such that there is sufficient shearing
occurring at the base of the convective boundary.  There is only a
slight increase in the growth rate for the \Fe\ mass around $t=18.5$
ms where the long-lived vortex first appears.  This is due to the
fact, mentioned above, that as the system evolves it goes from many
small vortices to a few large, dominant vortices.  It is only when the
circulation pattern of a particular vortex is large enough to strongly
interact with the lower convection boundary that we get the shearing
and enrichment of the convective region; this occurs around $t \sim 5$
ms.  The addition of \Fe\ to the convective region has a small but
noticeable effect on the conductivity; for example, a displacement of
$\sim 1\%$ \He\ for \Fe\ near the base of the accreted layer, with all
other things being constant, gives a $\sim 4\%$ decrease in
conductivity.  This could play an important role in adjusting the
flame speed once ignited.

%% \begin{figure*}[th]
%%   \begin{center}
%%     \plotone{c12_early}
%%   \end{center}
%%   \caption{\label{Fig:c12 early} Colormap plot of \C\ mass fraction
%%     with velocity vectors for the same region and times as shown in
%%     Figure \ref{Fig:early adiabatic excess evolution}.}
%% \end{figure*}

%% \begin{figure*}[th]
%%   \begin{center}
%%     \plotone{c12_late}
%%   \end{center}
%%   \caption{\label{Fig:c12 late} Colormap plot of \C\ mass fraction
%%     with velocity vectors for the same region and times as shown in
%%     Figure \ref{Fig:late adiabatic excess evolution}.}
%% \end{figure*}

%% \begin{figure*}[th]
%%   \begin{center}
%%     \plotone{Fe_enrichment}
%%   \end{center}
%%   \caption{\label{Fig:iron enrichment} Plots showing the
%%     \Fe\ enrichment of the convective region.  The left panel shows
%%     the evolution of the average \Fe\ mass fraction starting from the
%%     initial model distribution (solid thick line) and ending after
%%     $30$ ms of evolution (dashed line).  The right panel shows the
%%     total mass of \Fe\ in the convective region as a function of time.
%%     Note the log scale of the horizontal axis in the right plot.}
%% \end{figure*}

Figure \ref{Fig:long enucdot} shows the evolution of the maximum value
of $\Hnuc$ throughout the duration of the \cold\ model simulation.  The
inset plot shows the early adjustment phase mentioned above.  The
initial jump in $\Hnuc$ is due to the fact that there is no well
established flow field that can efficiently advect away the energy
released from reactions.  Once the convective zone is well
established, the energy generation rate relaxes before making its
steady climb.  We note that we have not yet achieved runaway---the
rise in energy generation rate is still linear.  This climb is
temporarily interrupted by a couple of spikes similar to those seen in
Figure \ref{Fig:deltap spike closeup} when fresh fuel is advected to a
relatively hot region and burned quickly.  Although well organized at
later times, the convective fluid flow is slow with respect to the
sound speed.  Figure \ref{Fig:Mach number} shows the maximum Mach
number in the computational domain as a function of time; this value
never exceeds $0.08$ in our simulation.  The average value of the Mach
number in the convective region rarely exceeds $\sim 0.02$ during our
$30$ ms simulation.

%% \begin{figure*}[th]
%%   \begin{center}
%%     \plotone{long_enucdot}
%%   \end{center}
%%   \caption{\label{Fig:long enucdot} Plot of the maximum $\Hnuc$ in the
%%     \cold\ model simulation as a function of time.  The inset plot
%%     shows the early adjustment phase associated with Figures
%%     \ref{SubFig:ad excess 0.4} and \ref{SubFig:ad excess 0.8}.  The
%%     spikes are similar to those seen in the bottom panel of Figure
%%     \ref{Fig:deltap spike closeup}.}
%% \end{figure*}

%% \begin{figure*}[th]
%%   \begin{center}
%%     \plotone{Machno}
%%   \end{center}
%%   \caption{\label{Fig:Mach number}Plot of the maximum Mach number in
%%     the \cold\ model simulation as a function of time.  The slow
%%     convective flow justifies the use of a low Mach number
%%     approximation method.}
%% \end{figure*}

%==========================================================================
% Conclusions
%==========================================================================
\section{Conclusions}\label{Sec:Conclusions}
We have described some of the challenges and important concepts to
keep in mind when performing multidimensional simulations of XRBs.
The major results can be summarized as follows:
\begin{itemize}
  \item To get a system that is much closer to thermal instability in
    multiple dimensions, the semi-analytic one-dimensional models
    should augment the local cooling rate estimate, (\ref{eq:approx
      cooling}), to include cooling due to convection.
  \item Properly resolving the burning layer using the initial models
    considered here requires a spatial resolution of $0.5$ cm
    zone$^{-1}$, which is an order of magnitude higher than what has
    been presented in the literature for multidimensional models
    \citep{Lin:2006}.  It should be noted that our initial models
    differ in the underlying neutron star's composition---their
    \C\ opposed to our \Fe---and their models were evolved in one
    dimension through several bursts before being mapped into multiple
    dimensions.
  \item Under-resolving the burning layer leads to dramatic convective
    undershoot and the burning tends to die out.
  \item At the early times simulated here, the inclusion of thermal
    diffusion has little effect on the maximum energy generation rate,
    but does perhaps affect the depth at which this maximum occurs.
  \item The {\tt MAESTRO} algorithm we use allows us to capture the
    expansion of the atmosphere due to heating, which will be important
    in the modeling of PRE burst sources.
 \item The average thermal gradient in the convective region is nearly
   adiabatic but there are localized pockets and filamentary
   structures that are either super- or sub-adiabatic.
 \item The strong convection interacts with and churns up the
   underlying neutron star material, which slightly alters the
   conductivity of the convective region.
\end{itemize}

The initial selection of a value to use for the volume discrepancy
term in our simulations was based on experience with other
applications.  As we showed in Section \ref{Sec:Effects of the Volume
  Discrepancy Term}, the value used for the long duration simulation
in this paper, $f=0.3$, may not the optimal choice for the XRB problem.
Further investigation is required to determine which factors affect
the appropriate value of $f$, and to determine if the spikes in the
drift of the thermodynamic pressure from the base state pressure are
simply numerical artifacts of a poor choice of $f$.

The width of the computational domain used in our simulations is
adequate for the early evolution of the system; the size of any
individual convective cell is initially small with respect to the
width of the domain.  As the system evolves and the convection becomes
more established, the cells grow in size.  The nature of vorticity in
two dimensions is such that the smaller vortices merge to form a
single vortex.  In our simulations the cells grow to become a
significant fraction of the domain width and the flow becomes
dominated by a single vortex that interacts with itself through the
periodic boundary conditions.  By selecting a wider computational
grid, we could delay the formation of a single, dominant vortex.
Ideally the computational domain should be several pressure
scale-heights wide so that we should form multiples of these
convective cells that dominate the flow for an extended period of time
before merging into a single vortex.  Given our strict resolution
requirements, such a setup was computationally infeasible.

We plan to further investigate some of the topics mentioned above in
future work while studying mixed H/He bursts.  In such bursts the
majority of the energy release comes from burning hydrogen; the
nuclear reaction rates involved in such burning are less temperature
sensitive than the $3$-$\alpha$ rate used in the current paper.  This
may allow for a relaxed resolution requirement for properly resolving
the burning layer because the energy generation rate profile should
not be as sharply peaked as we have seen in our studies.  This would
allow for longer time evolution, which may allow us to say something
about whether or not the convective zone extends all the way to the
photosphere near outburst.  We will also be able to simulate larger
domains where we could address the effects of domain size on the
long-term evolution of the convective region and its \Fe\ enrichment.
Furthermore, we will begin investigating the effects of unprecedented
three-dimensional simulations of the convection that precedes the
outburst in an XRB and compare its properties to our two-dimensional
studies.  All of these simulations will rely heavily on {\tt
  MAESTRO}'s new AMR capability to reduce computational cost for
long-term evolution.  We will also investigate the effects of using
different methods of calculating opacities as well as updated reaction
rates, such as the $3$-$\alpha$ rate of \cite{Ogata_etal09}.

\acknowledgements

We thank Frank Timmes for making his conductivity and equation of
state routines publicly available and Ed Brown for useful discussions
regarding the possibility of relaxed resolution requirements for H
burning.  We also thank Andrew Cumming for providing the initial
semi-analytic models and Stan Woosley for using the {\tt Kepler} code
to augment these models.  The work at Stony Brook was supported by a
DOE/Office of Nuclear Physics Outstanding Junior Investigator award,
grant No.\ DE-FG02-06ER41448, to Stony Brook.  The work at LBNL was
supported by the SciDAC Program of the DOE Office of Mathematics,
Information, and Computational Sciences under the U.S. Department of
Energy under contract No.\ DE-AC02-05CH11231. This research utilized
resources at the New York Center for Computational Sciences at Stony
Brook University/Brookhaven National Laboratory, which is supported by
the U.S. Department of Energy under Contract No. DE-AC02-98CH10886 and
by the State of New York.  This research used resources of the
National Energy Research Scientific Computing Center, which is
supported by the Office of Science of the U.S. Department of Energy
under Contract No. DE-AC02-05CH11231.

%==========================================================================
% FIGURES
%==========================================================================
\clearpage

\clearpage

\clearpage

\clearpage

\clearpage
\begin{figure*}[th]
\begin{center}
\plotone{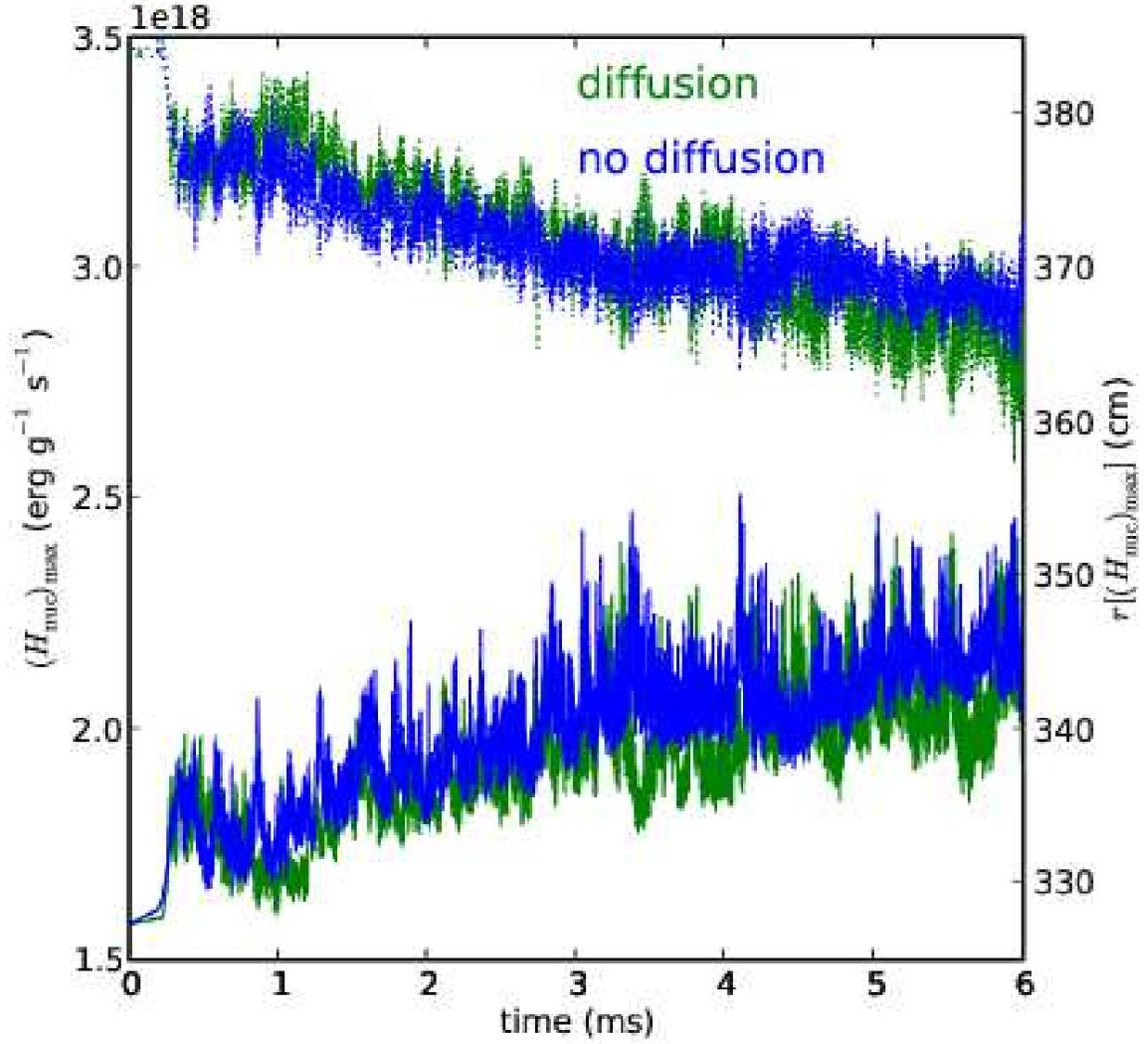}
\end{center}
\caption{\label{Fig:effect of diffusion}Evolution of $\mymax{\Hnuc}$
  (solid lines) and its vertical location (dashed lines) as a function
  of time for the \hot\ model both with (green) and without (blue)
  thermal diffusion.  }
\end{figure*}

\clearpage
\begin{figure*}[th]
\begin{center}
\plotone{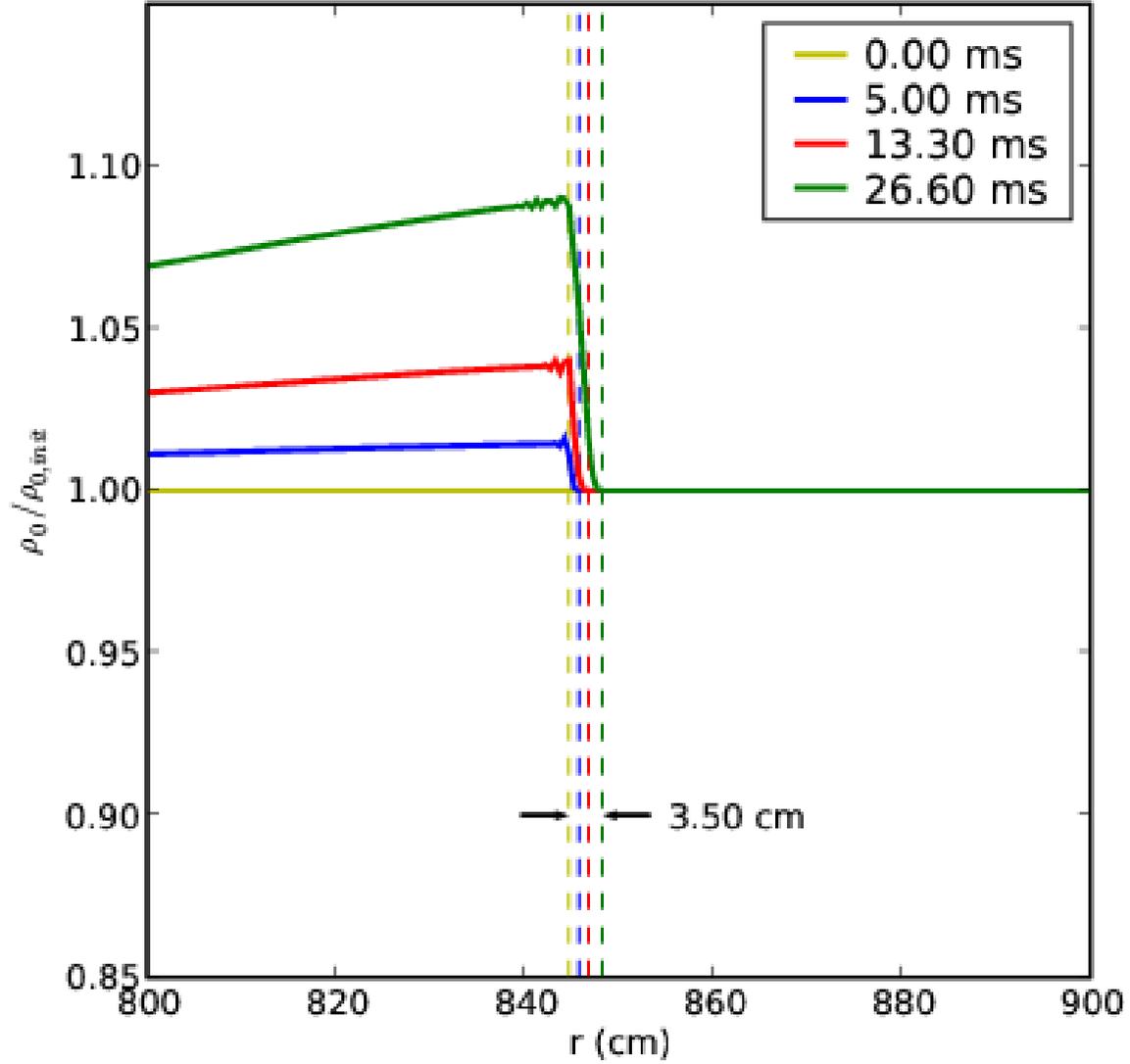}
\end{center}
\caption{\label{Fig:expansion of base state} Expansion of the base
  state due to heating.  Plotted is the ratio of base state density,
  $\rho_0$, to the initial ($t=0$) base state density,
  $\rho_{0,init}$, near the surface of the atmosphere for the
  \cold\ model.  We define the surface to be where $\rho = \rho_{\rm
    cutoff}$ and it is represented by the vertical dashed lines.  The
  base state has expanded $3.5$ cm in $26.6$ ms of evolution. }
\end{figure*}

\clearpage
\begin{figure*}[th]
\begin{center}
\epsscale{0.9}
\plotone{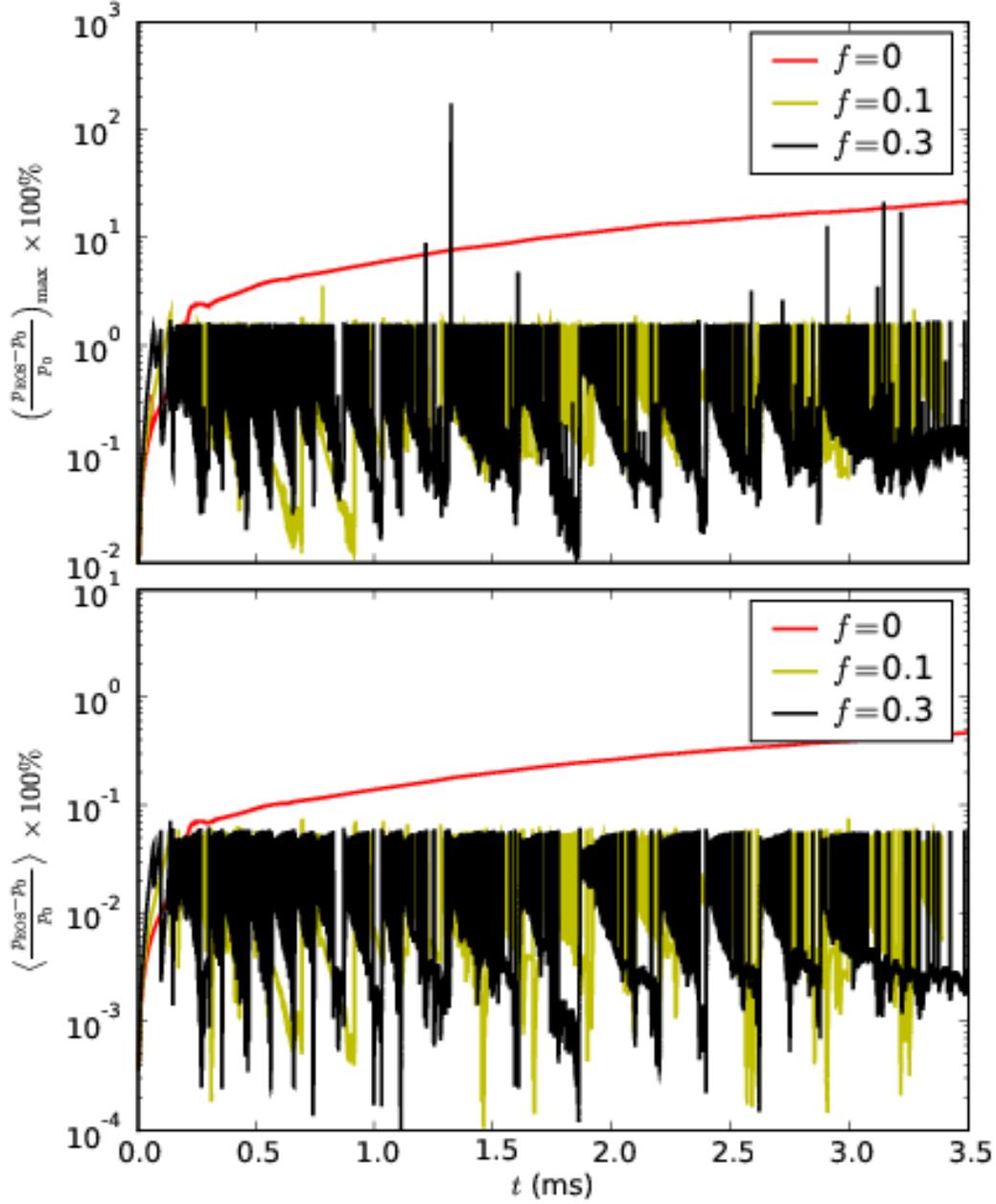}
\epsscale{1.0}
\end{center}
\caption{\label{Fig:volume discrepancy effect} Effects of the volume
  discrepancy factor as characterized by the percent difference
  between the thermodynamic pressure as given by the equation of state,
  $p_\text{EOS}$, and the base state pressure, $p_0$, for the
  \hot\ model.  The top panel shows the maximum value whereas the
  bottom panel shows the average value of the percent difference in the
  computational domain.  Note the different vertical scales between the
  two plots.}
\end{figure*}

\clearpage
\begin{figure*}[th]
\begin{center}
\epsscale{0.9}
\plotone{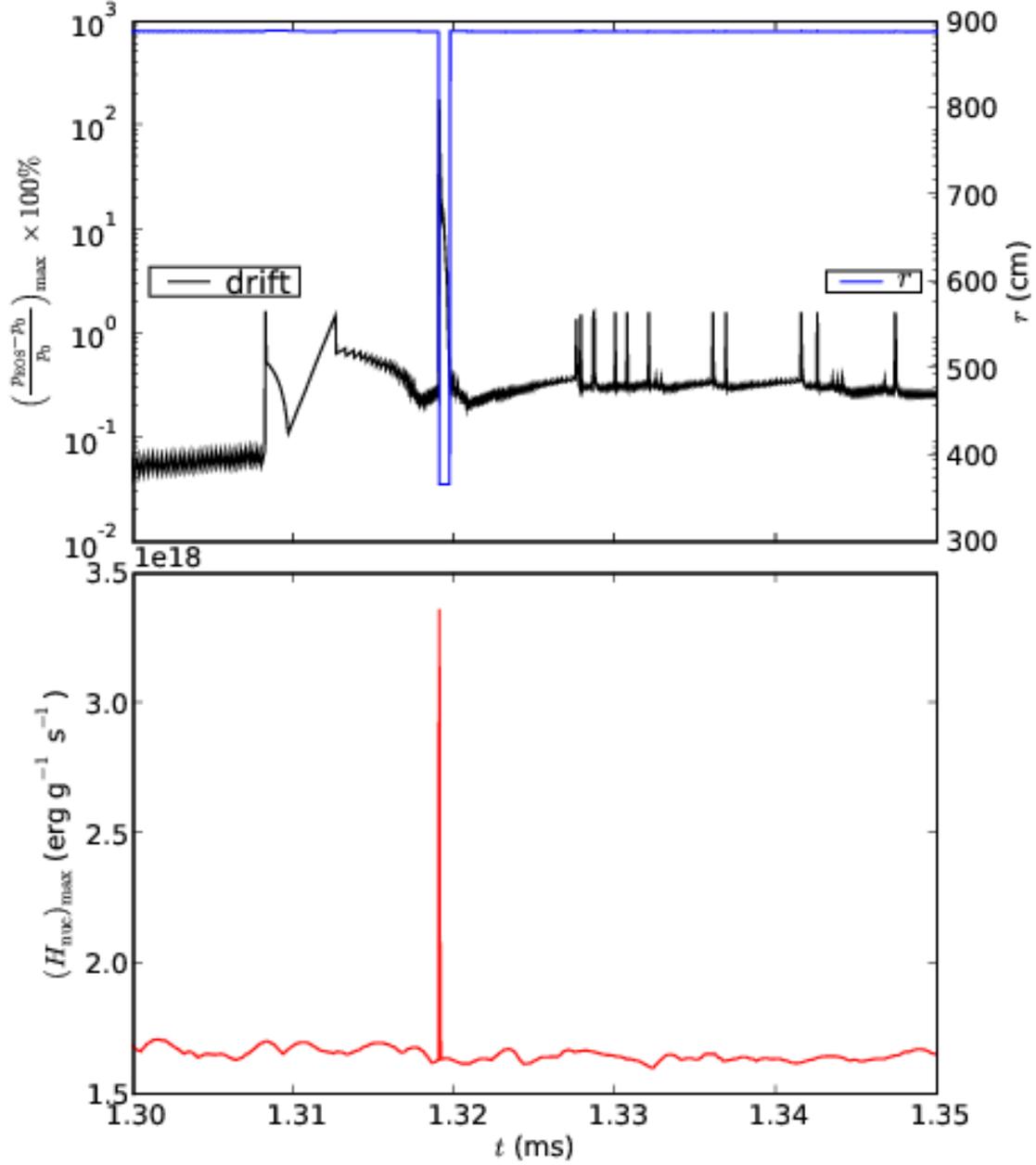}
\epsscale{1.0}
\end{center}
\caption{\label{Fig:deltap spike closeup} Closeup of the
  $\mathcal{O}(1)$ spike in the maximum value of the $f=0.3$ drift as
  seen in the top panel of Figure \ref{Fig:volume discrepancy
    effect}.  The top panel shows the drift value and its location in
  the domain; the bottom panel shows the maximum energy generation
  rate.  The large amount of energy released from the burning spike
  causes the thermodynamic pressure to differ from the hydrostatic
  base state pressure and therefore a spike in the drift.}
\end{figure*}

\clearpage
\begin{figure*}[th]
  \begin{center}
    \subfigure[$t = 0$ ms]{
      \epsscale{0.34}
      \plotone{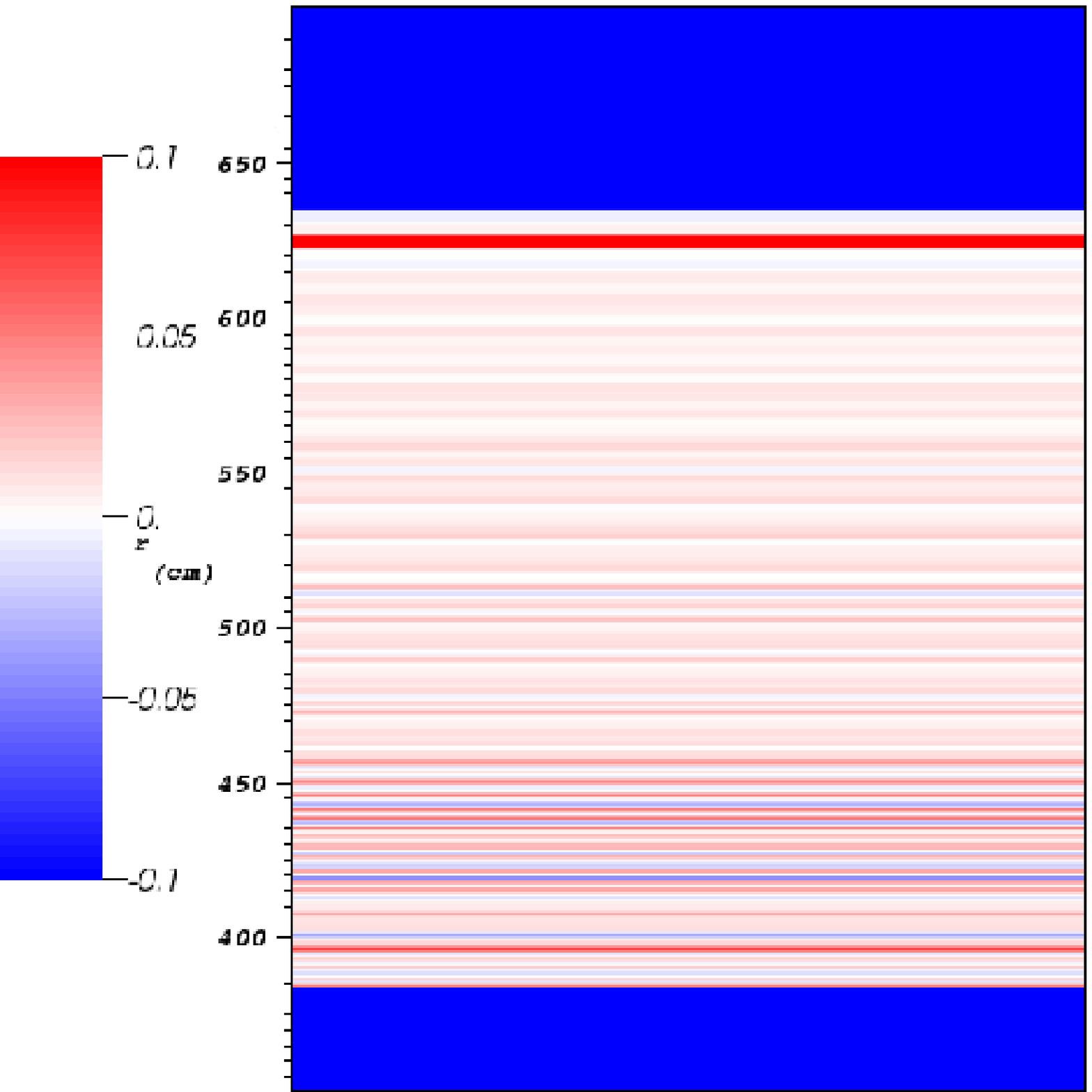}
      \epsscale{1.0}
      \label{SubFig:ad excess 0}
    }
    \subfigure[$t = 0.4$ ms]{
      \epsscale{0.25}
      \plotone{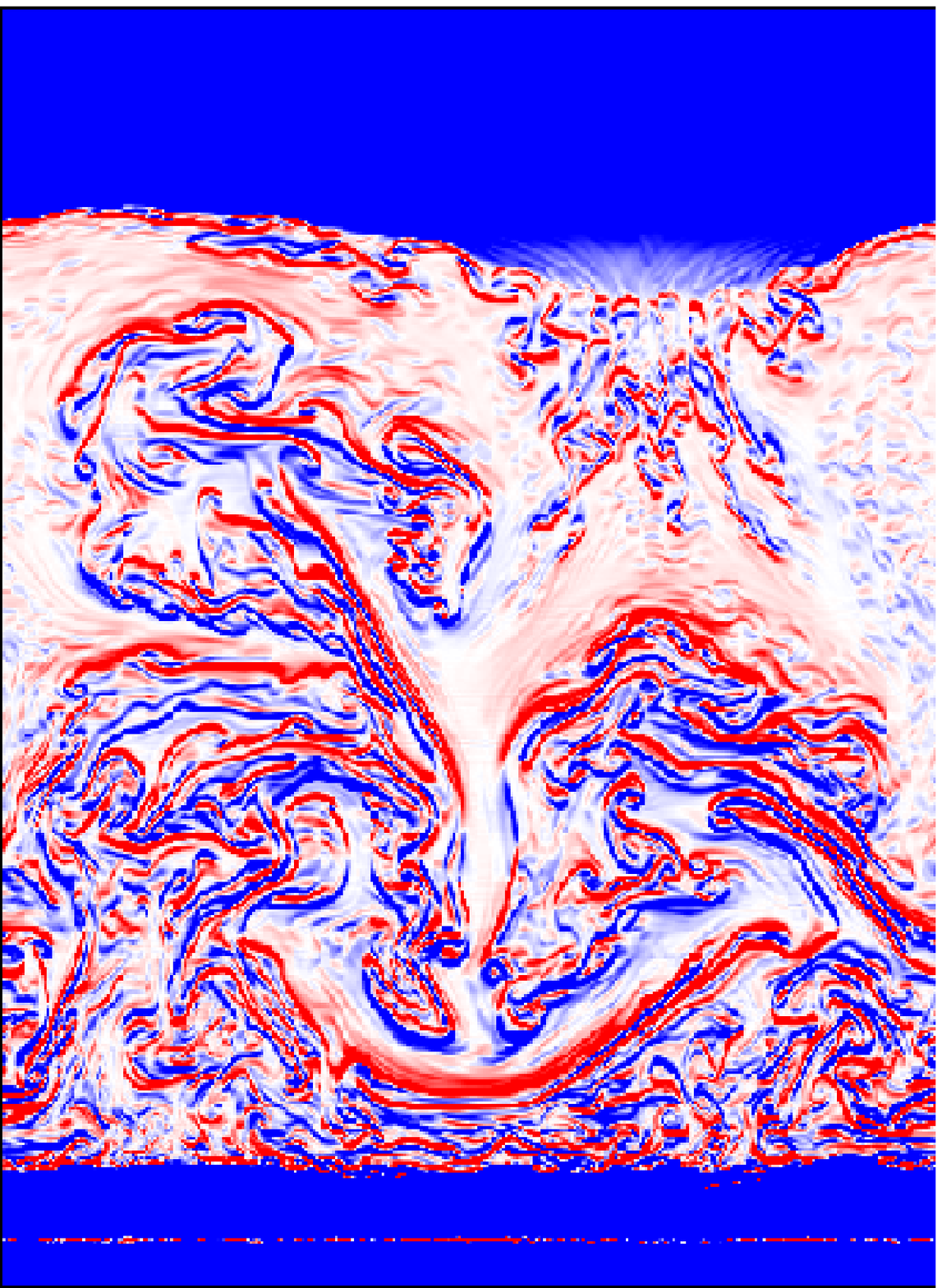}
      \epsscale{1.0}
      \label{SubFig:ad excess 0.4}
    }
    \subfigure[$t = 0.8$ ms]{
      \epsscale{0.25}
      \plotone{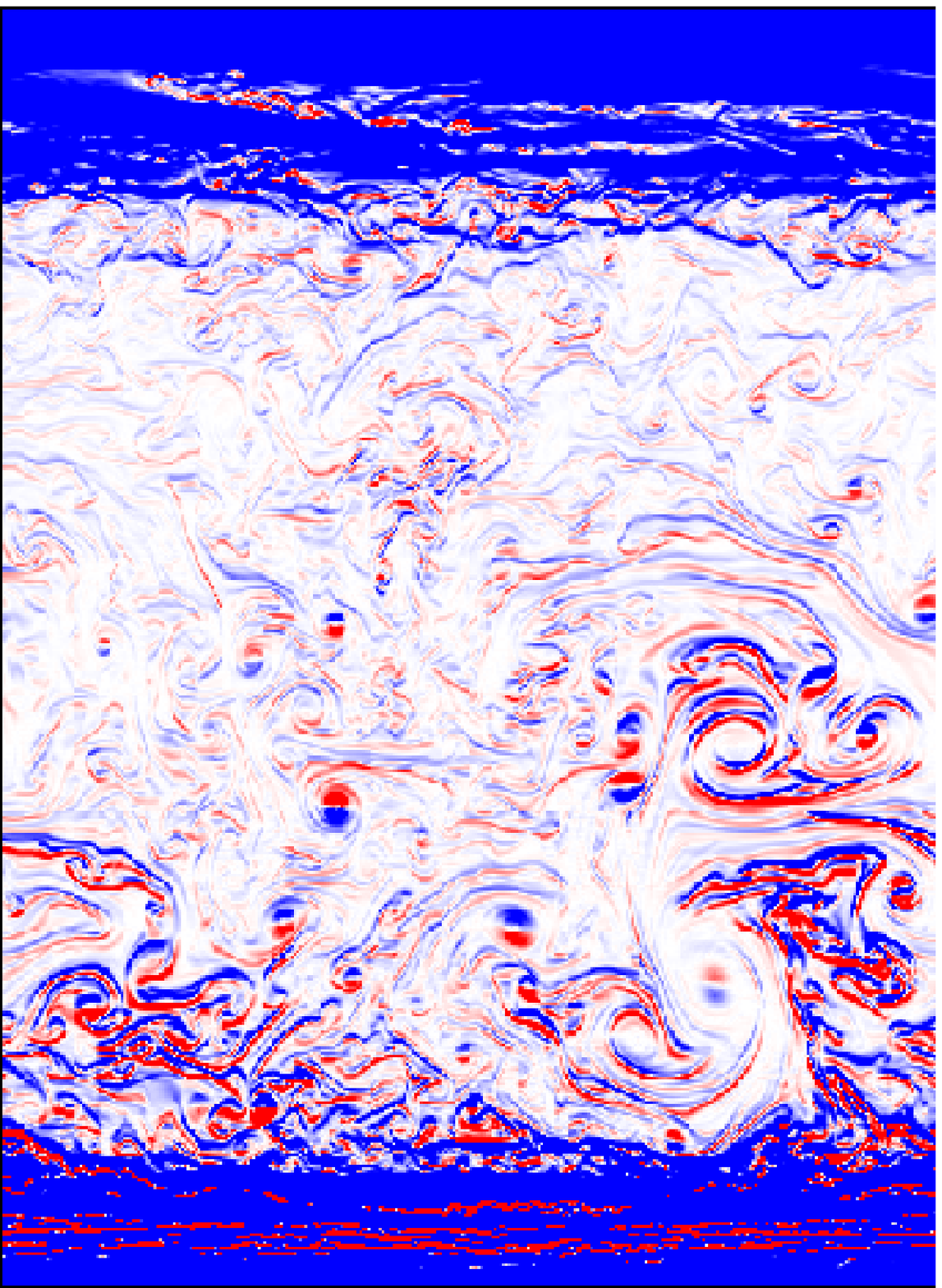}
      \epsscale{1.0}
      \label{SubFig:ad excess 0.8}
    }
    \subfigure[$t = 5$ ms]{
      \epsscale{0.34}
      \plotone{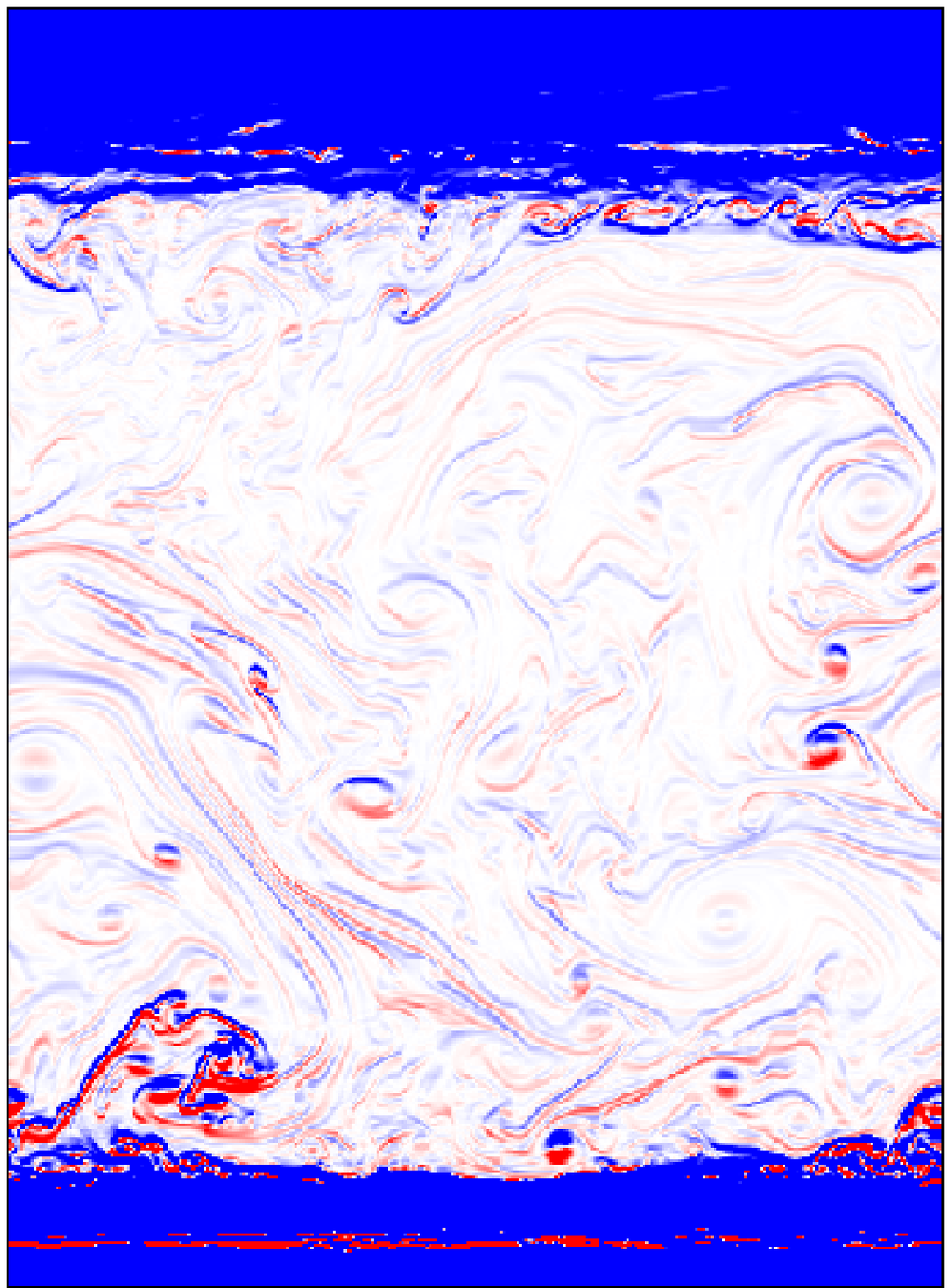}
      \epsscale{1.0}
      \label{SubFig:ad excess 5}
    }
    \subfigure[$t = 7.5$ ms]{
      \epsscale{0.25}
      \plotone{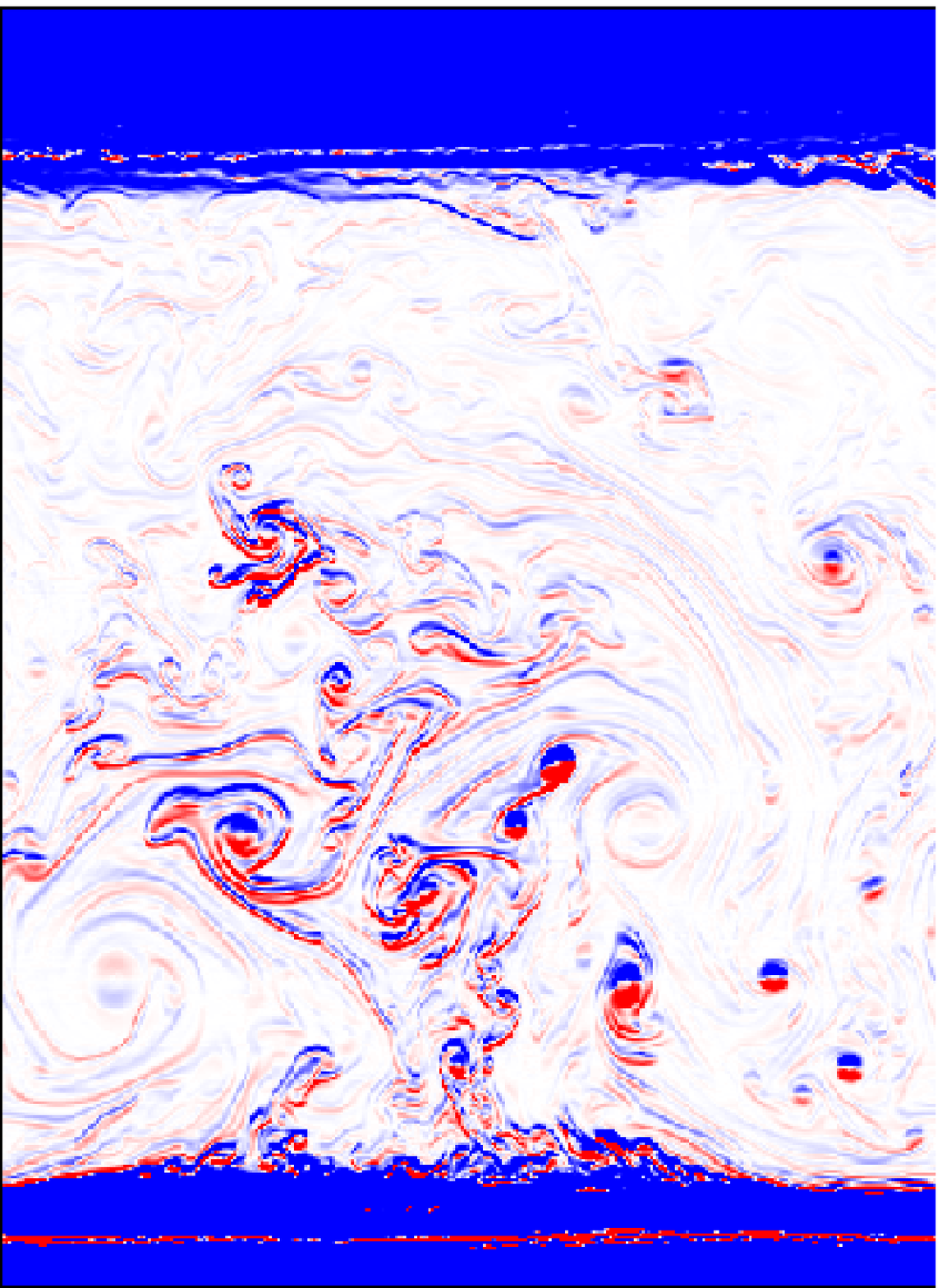}
      \epsscale{1.0}
      \label{SubFig:ad excess 7.5}
    }
    \subfigure[$t = 10$ ms]{
      \epsscale{0.25}
      \plotone{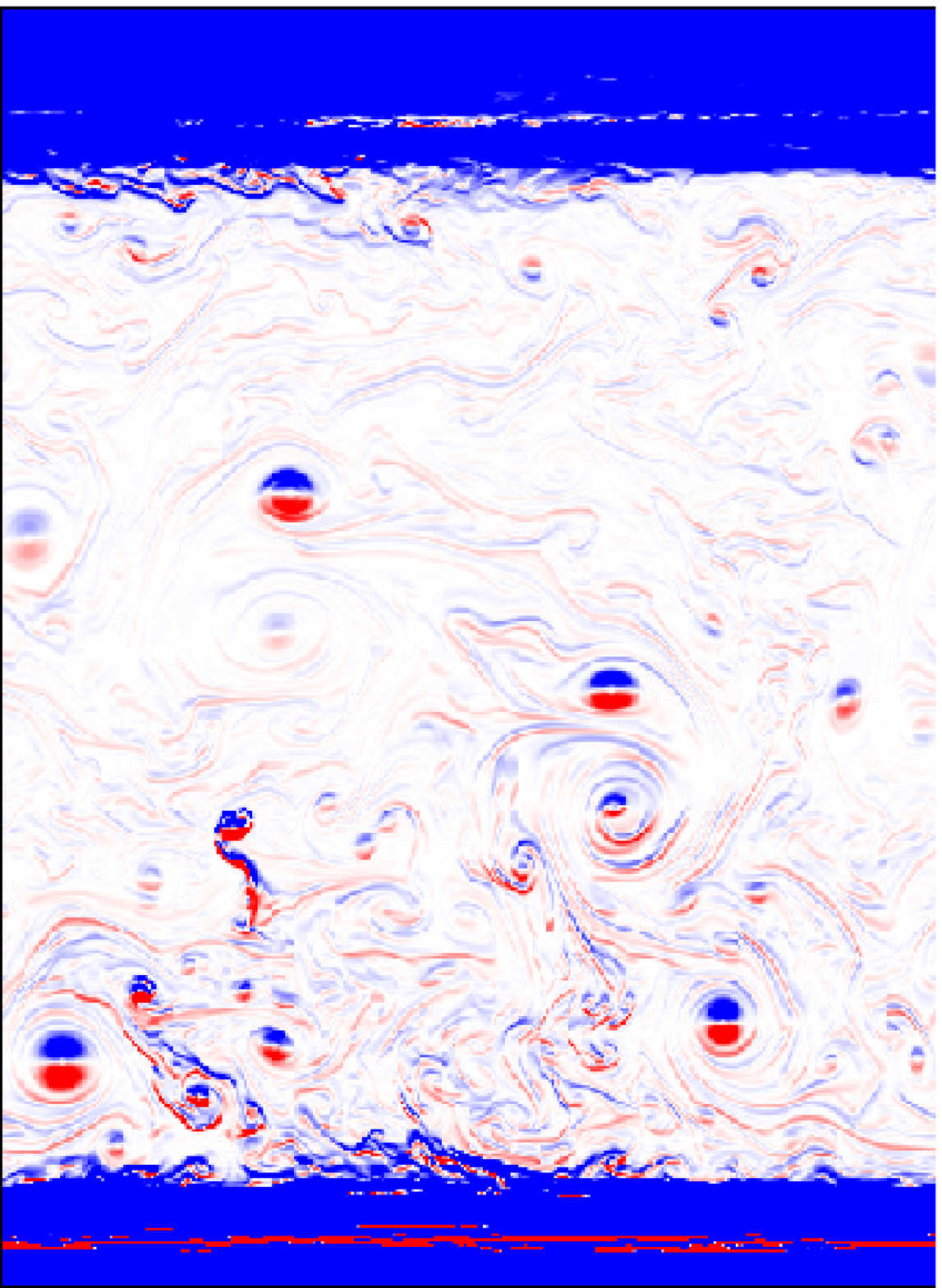}
      \epsscale{1.0}
      \label{SubFig:ad excess 10}
    }
  \end{center}
  \caption{\label{Fig:early adiabatic excess evolution} Colormap plot
    of the evolution of the adiabatic excess, $\Delta\nabla$, in the
    convective region for the \cold\ model.}
\end{figure*}

\clearpage
\begin{figure*}[th]
  \begin{center}
    \subfigure[$t = 18.5$ ms]{
      \epsscale{0.3}
      \plotone{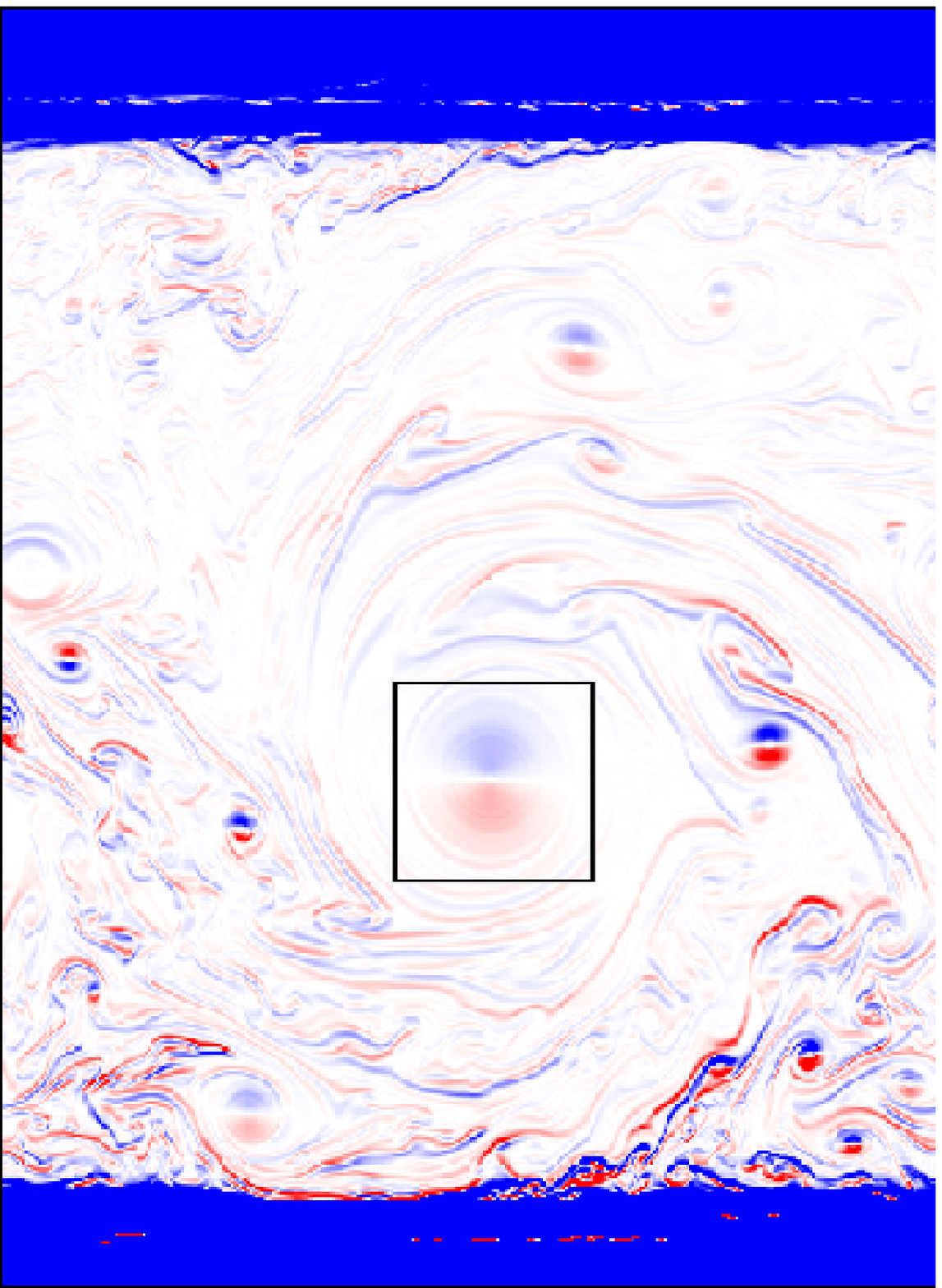}
      \label{SubFig:ad excess 18.5}
    }
    \subfigure[$t = 20.5$ ms]{
      \plotone{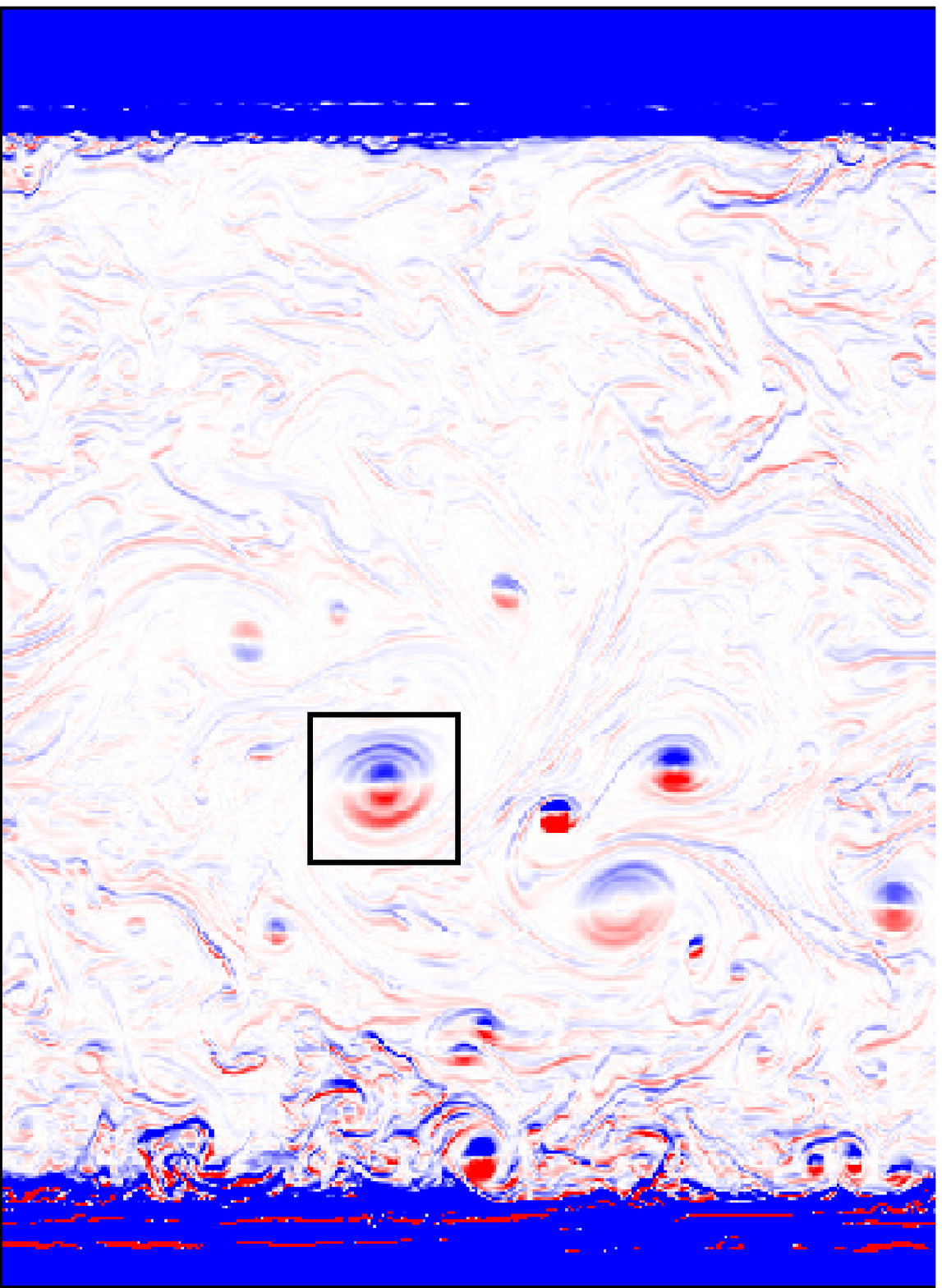}
      \label{SubFig:ad excess 20.5}
    }
    \subfigure[$t = 23$ ms]{
      \plotone{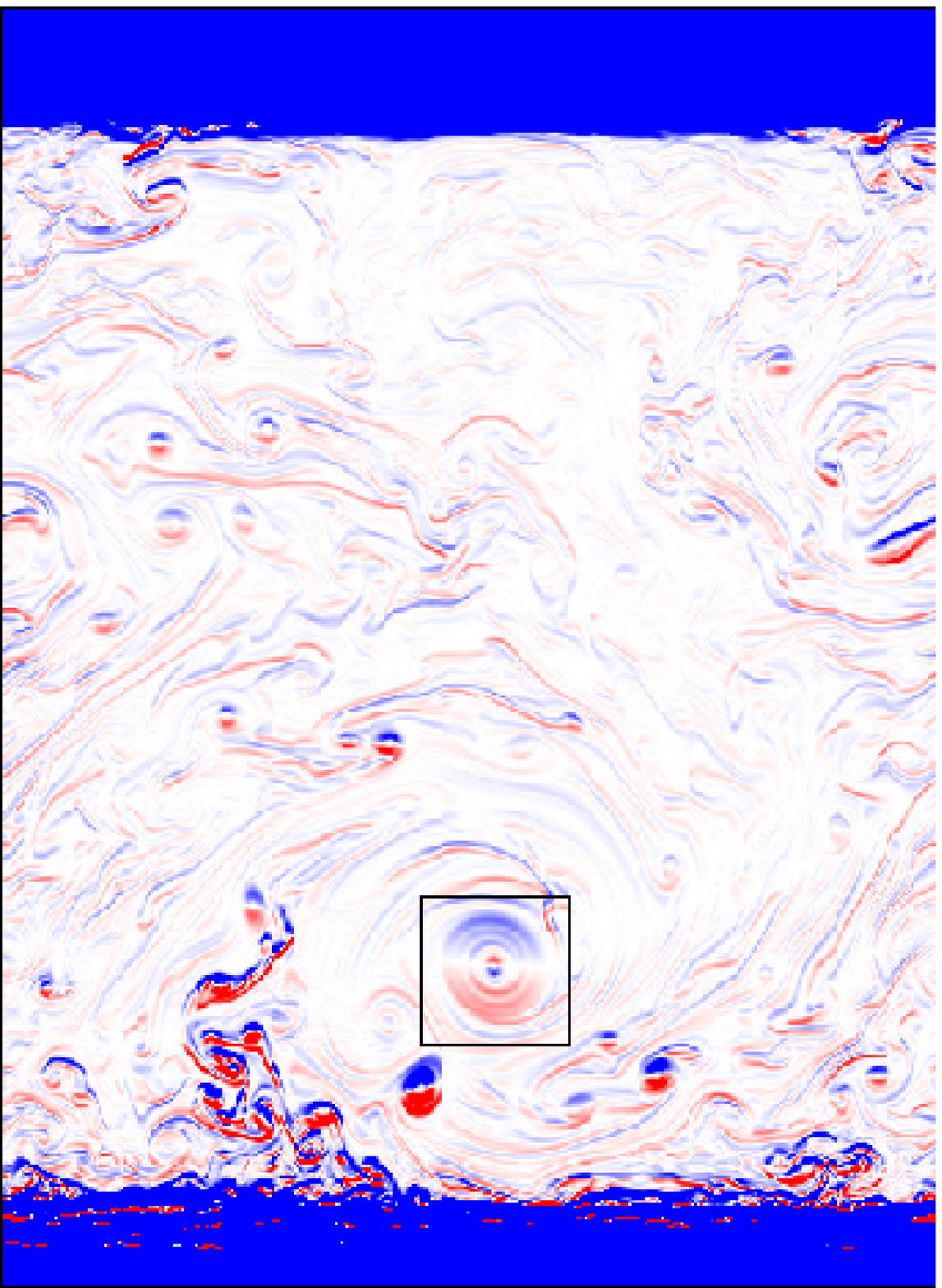}
      \label{SubFig:ad excess 23}
    }
    \subfigure[$t = 25$ ms]{
      \plotone{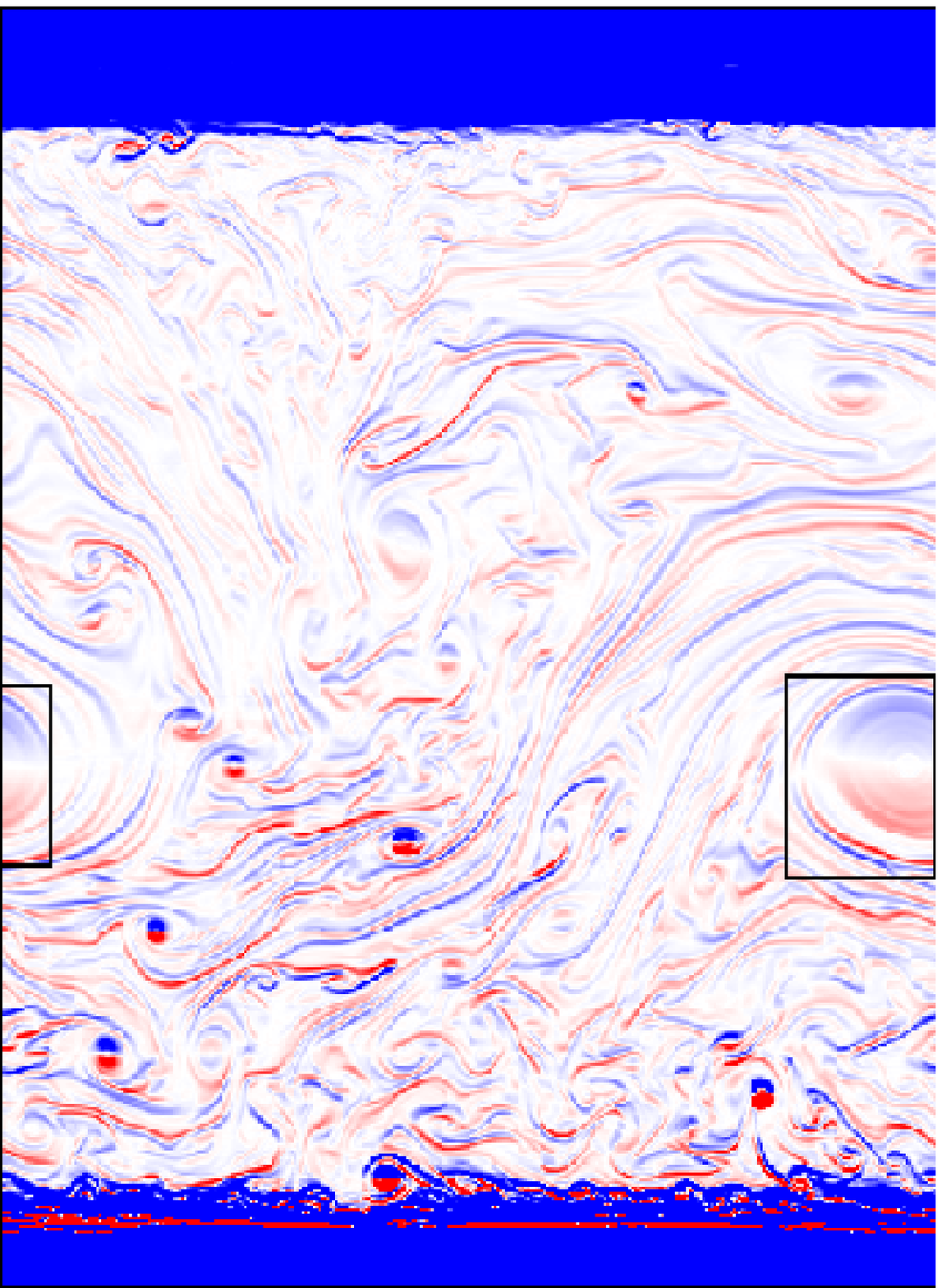}
      \label{SubFig:ad excess 25}
    }
    \subfigure[$t = 26$ ms]{
      \plotone{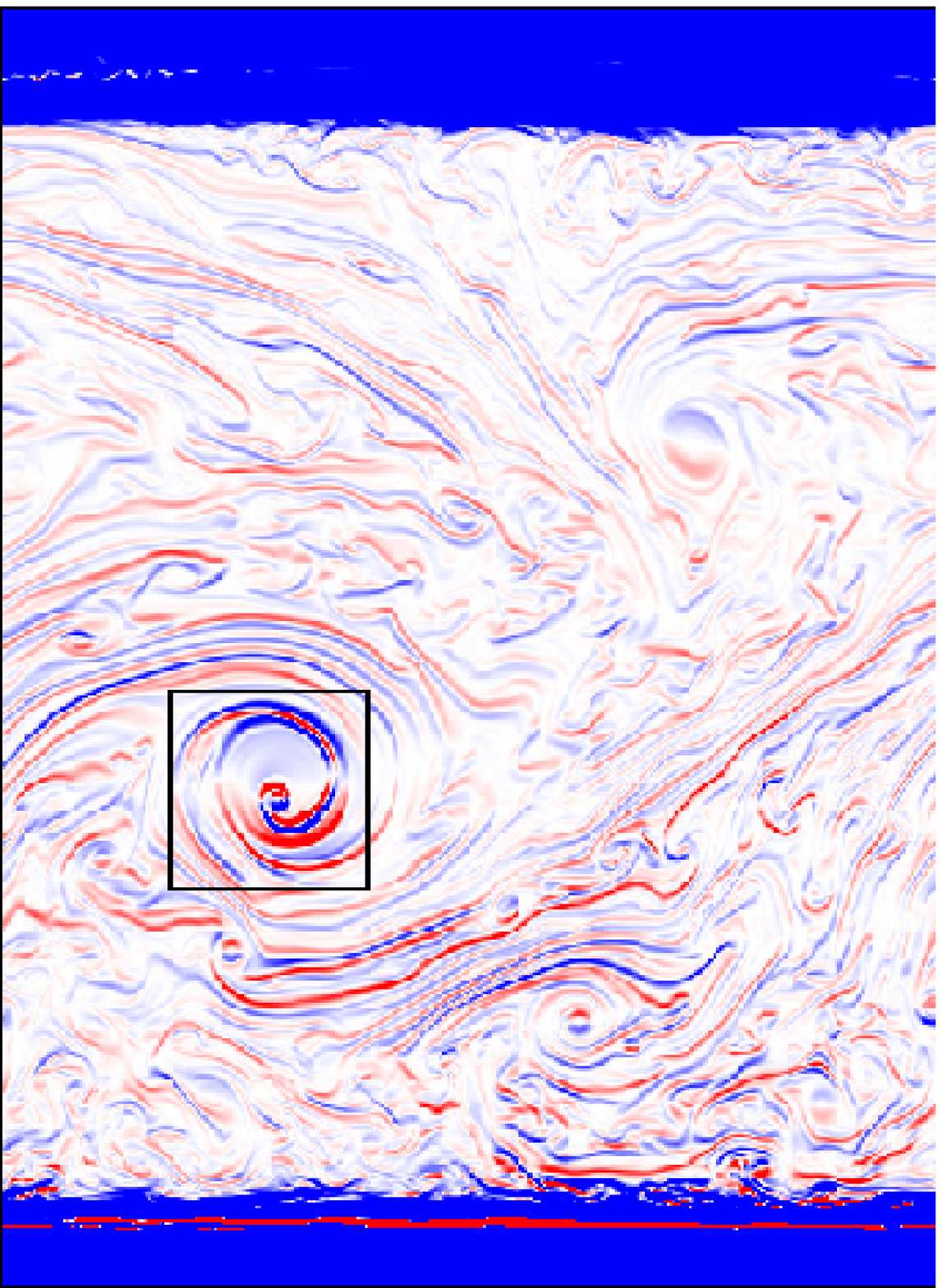}
      \label{SubFig:ad excess 26}
    }
    \subfigure[$t = 28$ ms]{
      \plotone{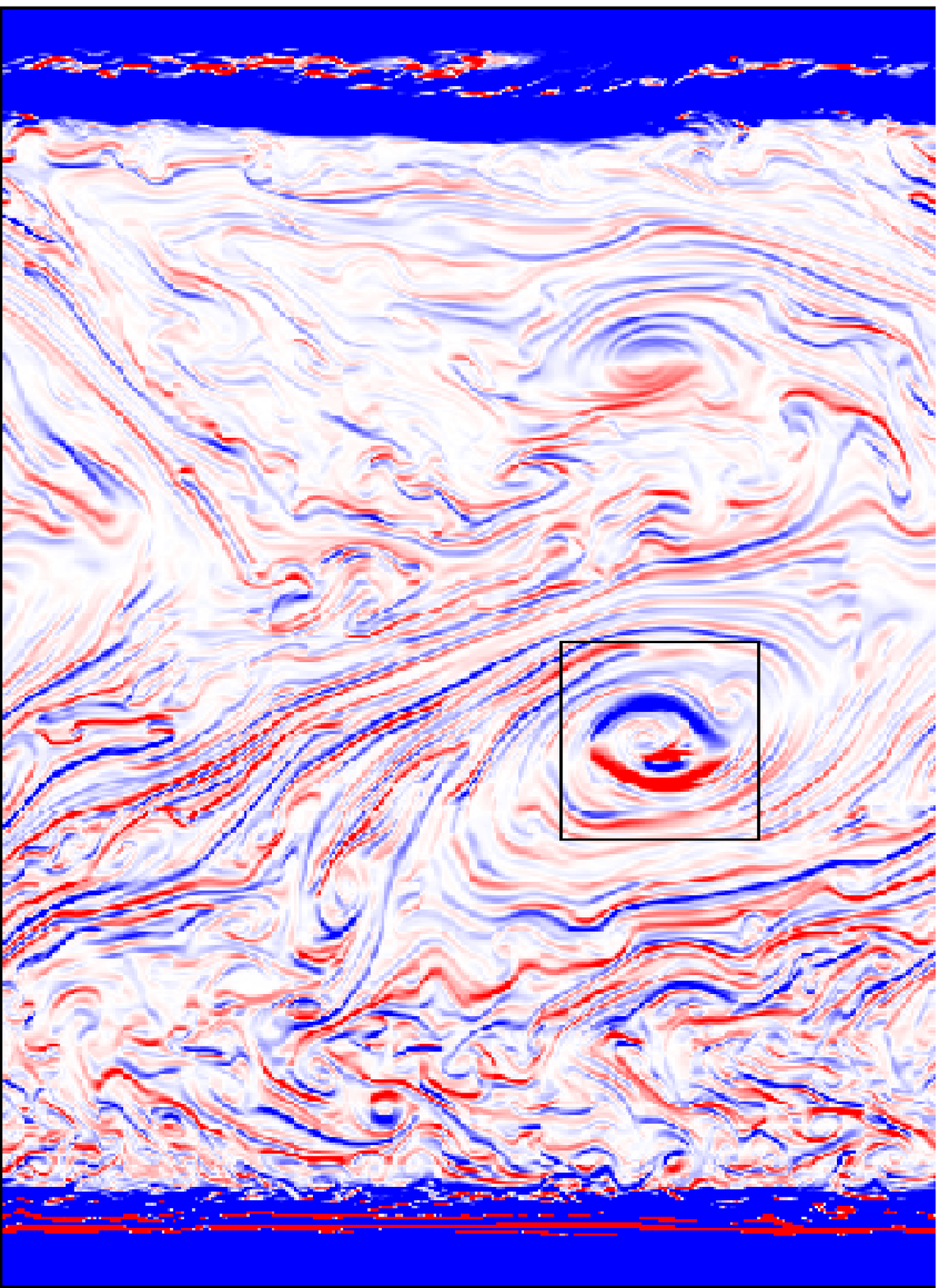}
      \epsscale{1.0}
      \label{SubFig:ad excess 28}
    }
  \end{center}
  \caption{\label{Fig:late adiabatic excess evolution} Same as Figure
    \ref{Fig:early adiabatic excess evolution} but at later times.
    The boxes show the location of a single feature that, once formed,
    lasts for the remainder of the simulation.}
\end{figure*}

\clearpage
\begin{figure*}[th]
  \centering
  \subfigure[Example Convective Profiles]{
    \plotone{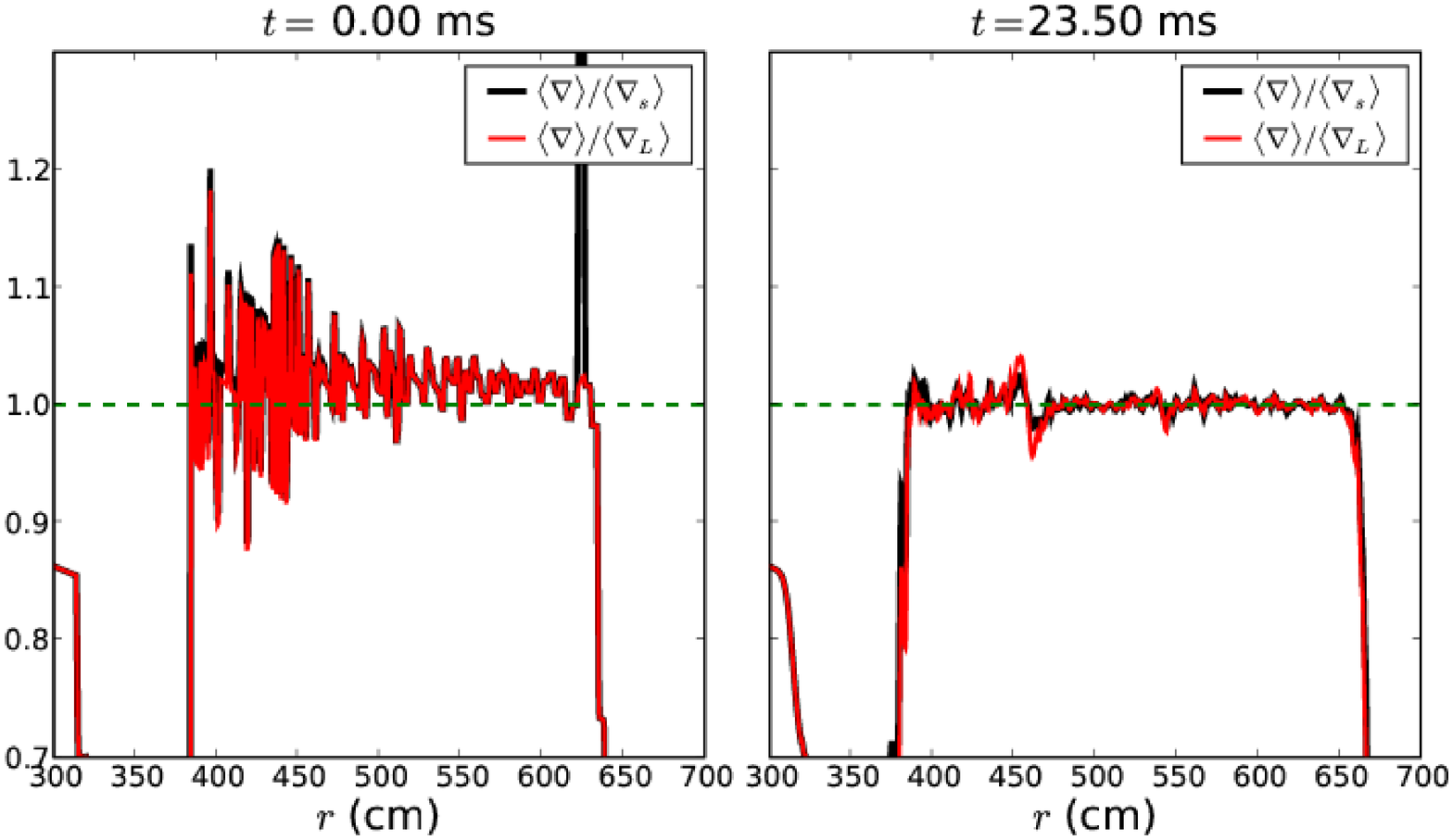}
    \label{SubFig:example convective profiles}
  }
  \subfigure[Convective Region Extent]{
    \plotone{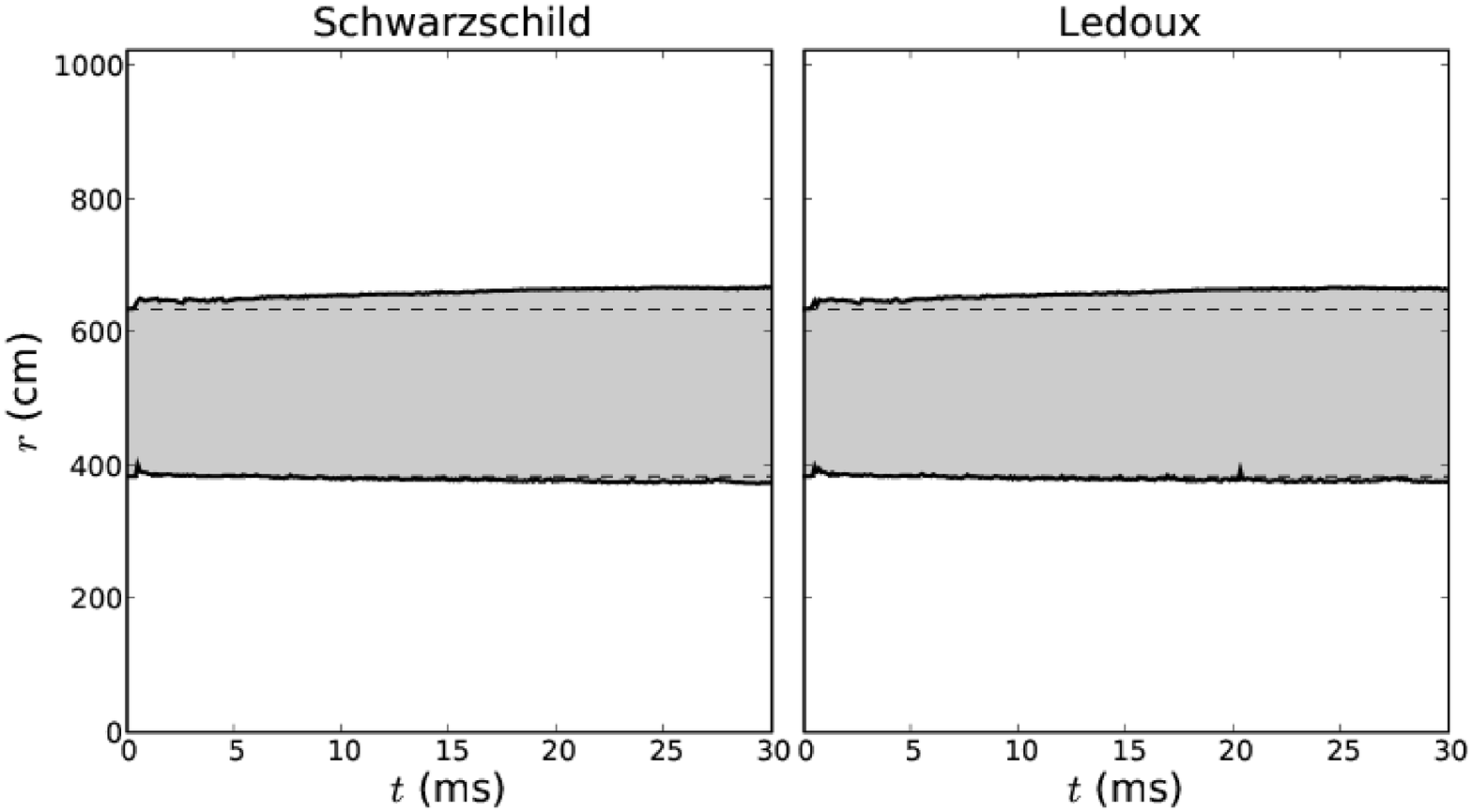}
    \label{SubFig:extents of convection}
  }
  \caption{\label{Fig:convective boundaries}Analysis of the extent of
    the convective region.  Panel (a) shows the convective profiles
    for both the Schwarzschild and Ledoux instability criteria at two
    different times.  Panel (b) shows the extent of the convective
    region as a function of time as determined by both instability
    criteria.}
\end{figure*}

\clearpage
\begin{figure*}[th]
  \begin{center}
    \plotone{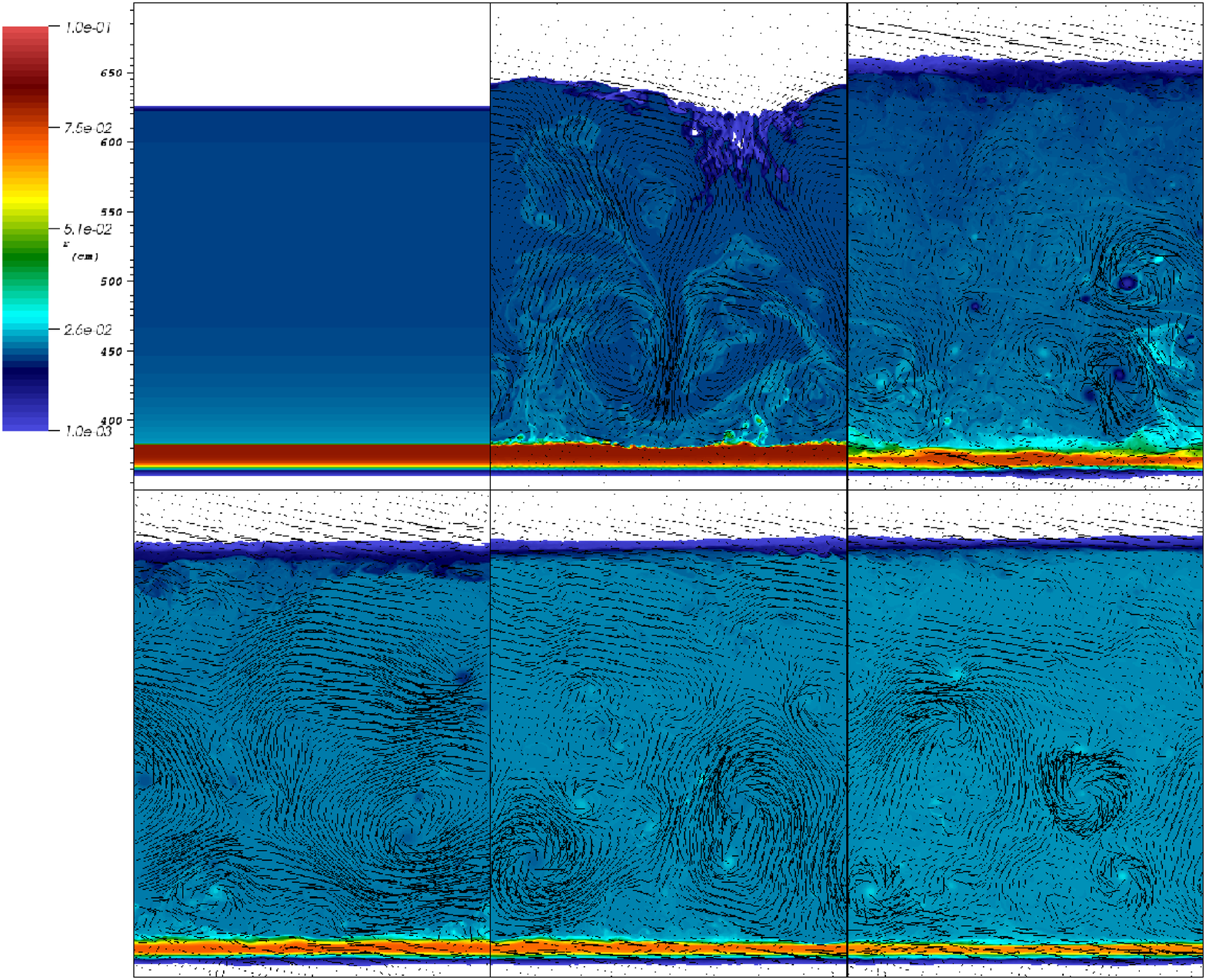}
  \end{center}
  \caption{\label{Fig:c12 early} Colormap plot of \C\ mass fraction
    with velocity vectors for the same region and times as shown in
    Figure \ref{Fig:early adiabatic excess evolution}.}
\end{figure*}

\clearpage
\begin{figure*}[th]
  \begin{center}
    \plotone{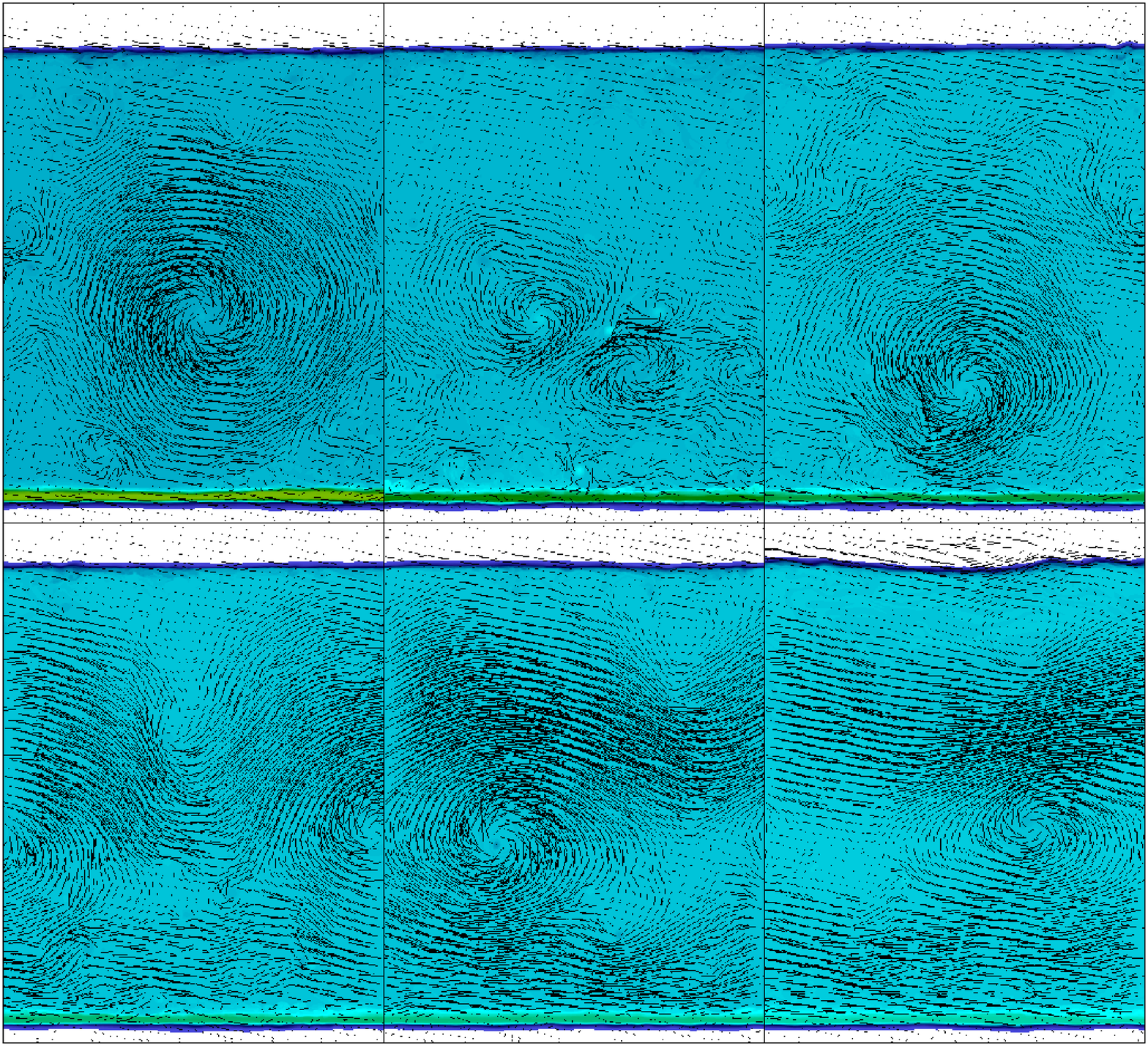}
  \end{center}
  \caption{\label{Fig:c12 late} Colormap plot of \C\ mass fraction
    with velocity vectors for the same region and times as shown in
    Figure \ref{Fig:late adiabatic excess evolution}.}
\end{figure*}

\clearpage
\begin{figure*}[th]
  \begin{center}
    \plotone{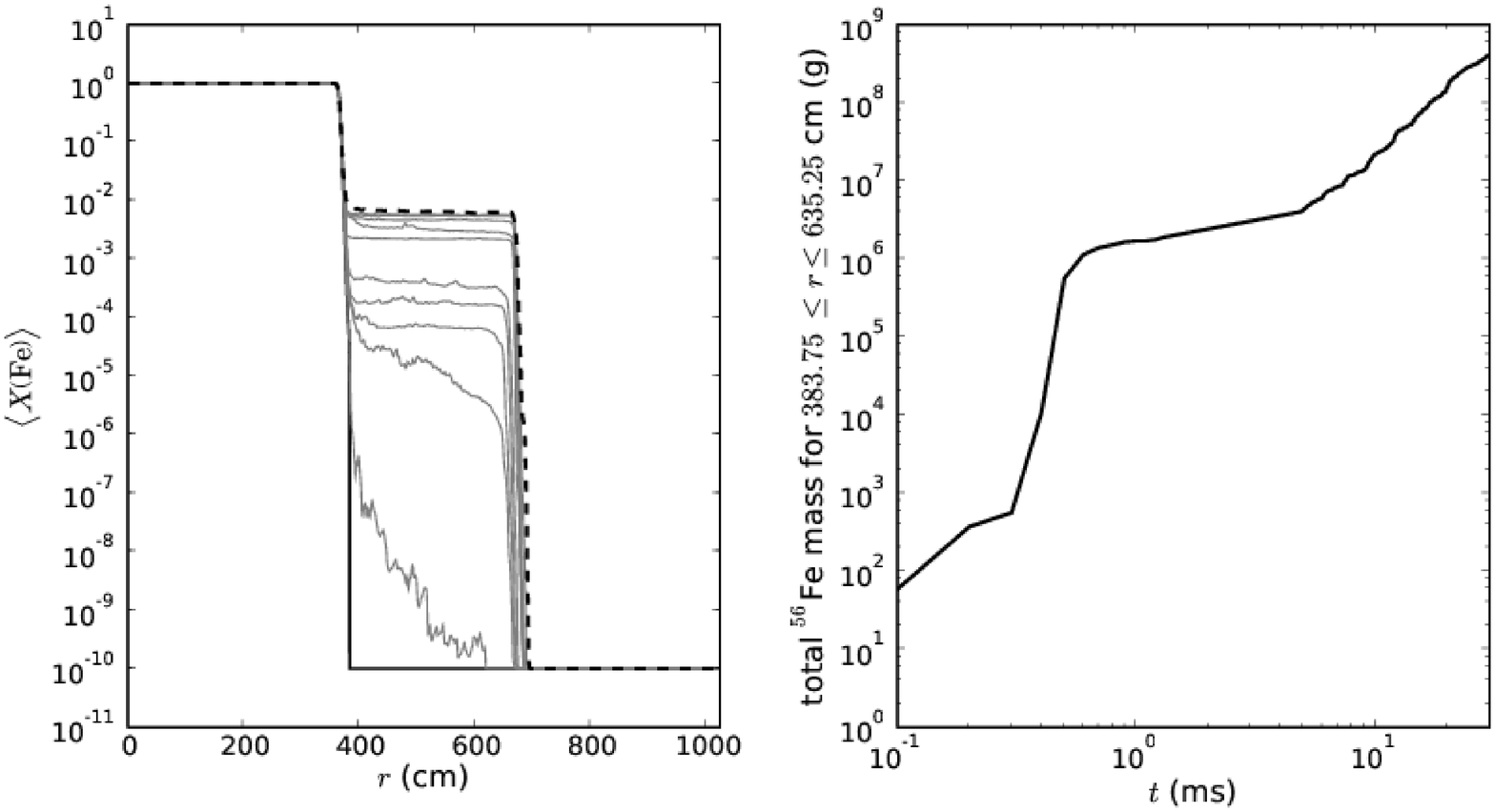}
  \end{center}
  \caption{\label{Fig:iron enrichment} Plots showing the
    \Fe\ enrichment of the convective region.  The left panel shows
    the evolution of the average \Fe\ mass fraction starting from the
    initial model distribution (solid thick line) and ending after
    $30$ ms of evolution (dashed line); the thin grey lines show the
    evolution at the intermediate times shown in Figures
    \ref{Fig:early adiabatic excess evolution} and \ref{Fig:late
      adiabatic excess evolution}.  The right panel shows the total
    mass of \Fe\ in the convective region as a function of time.  Note
    the log scale of the horizontal axis in the right plot.}
\end{figure*}

\clearpage
\begin{figure*}[th]
  \begin{center}
    \plotone{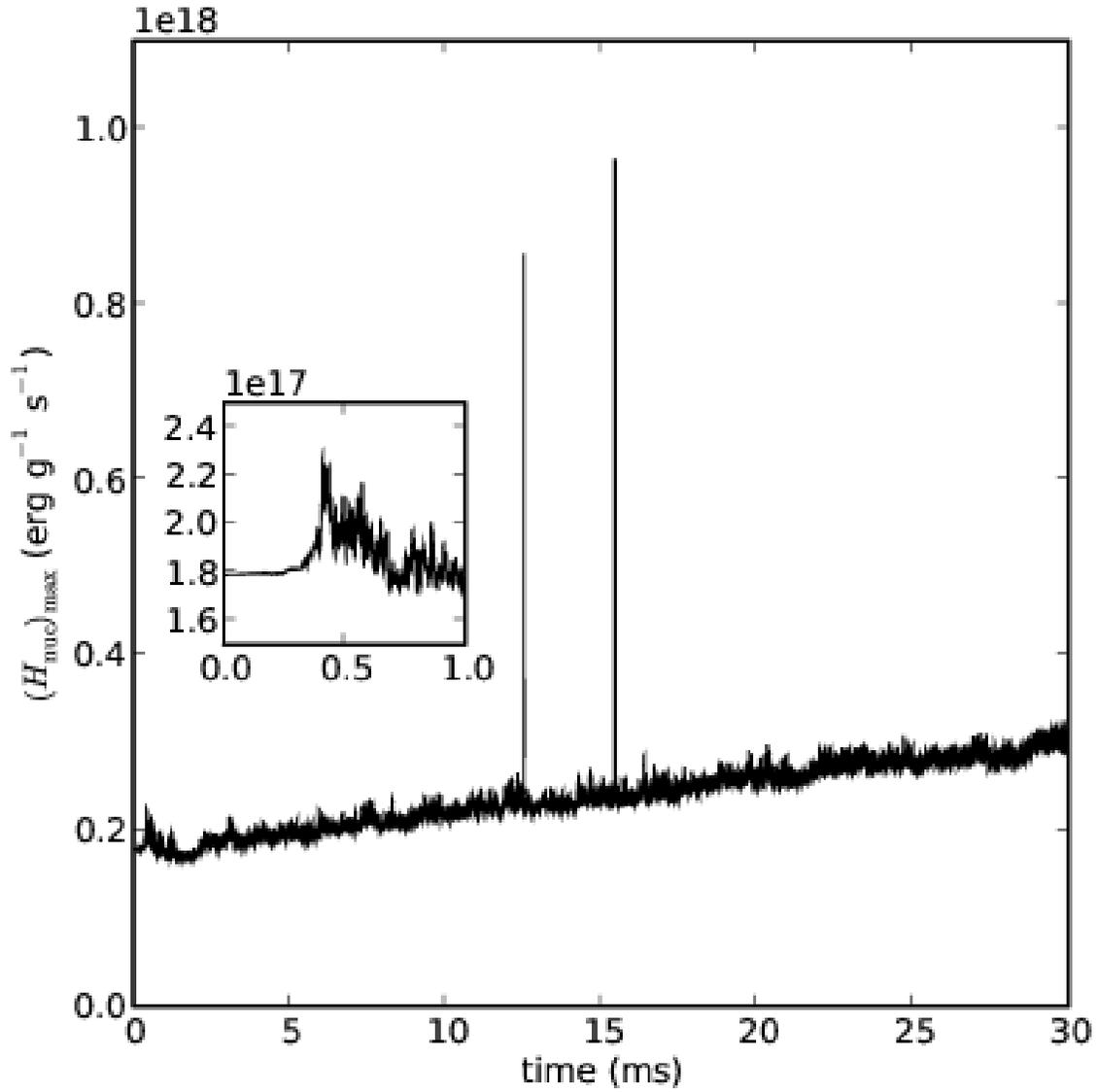}
  \end{center}
  \caption{\label{Fig:long enucdot} Plot of the maximum $\Hnuc$ in the
    \cold\ model simulation as a function of time.  The inset plot
    shows the early adjustment phase associated with Figures
    \ref{SubFig:ad excess 0.4} and \ref{SubFig:ad excess 0.8}.  The
    spikes are similar to that seen in the bottom panel of Figure
    \ref{Fig:deltap spike closeup}, and are caused by the rapid
    burning of fresh fuel as it is brought into the burning layer by
    the turbulent convection.}
\end{figure*}

\clearpage
\begin{figure*}[th]
  \begin{center}
    \plotone{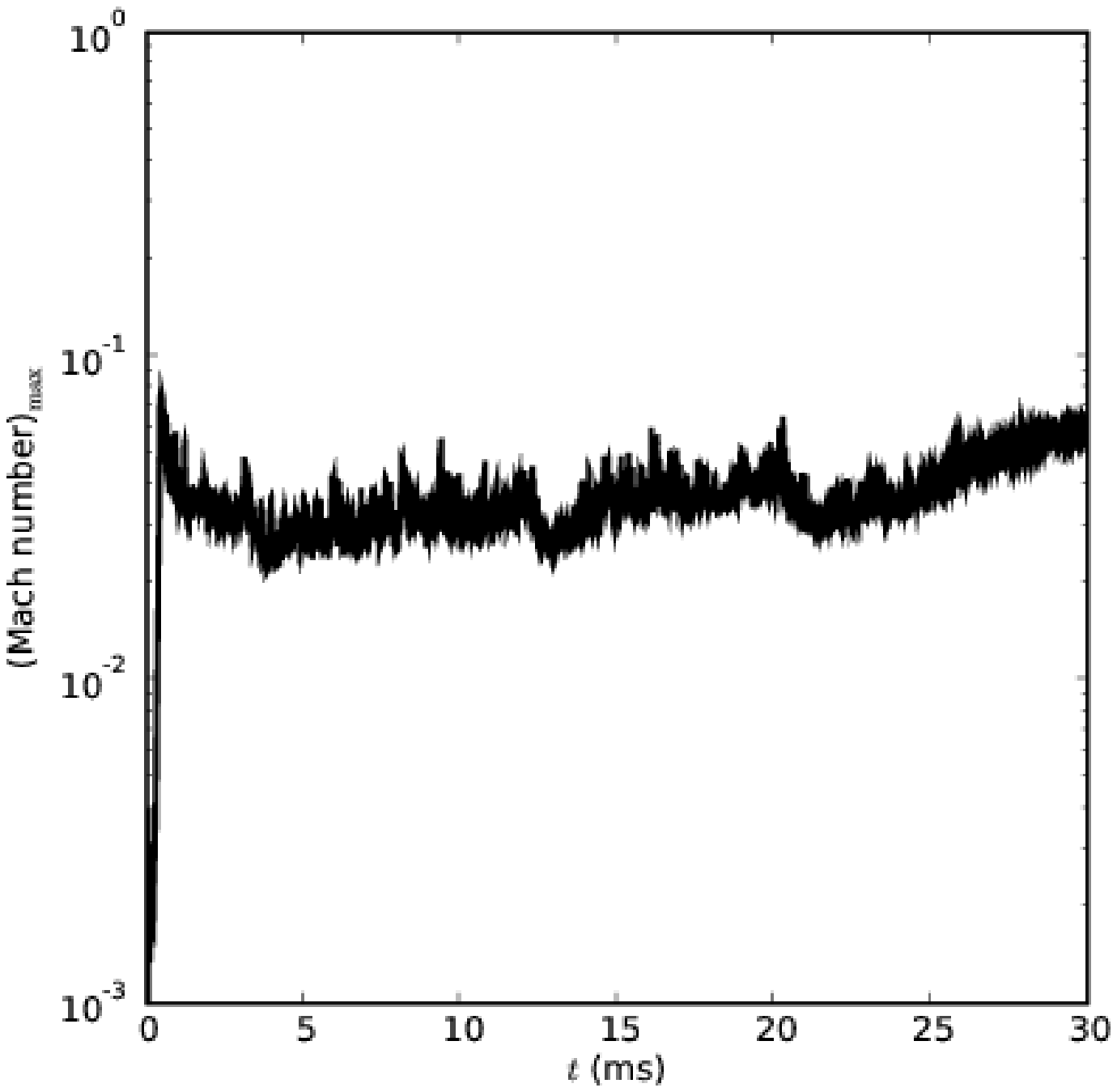}
  \end{center}
  \caption{\label{Fig:Mach number}Plot of the maximum Mach number in
    the \cold\ model simulation as a function of time.  The slow
    convective flow justifies the use of a low Mach number
    approximation method.}
\end{figure*}

%==========================================================================
% References
%==========================================================================
\clearpage
%\bibliographystyle{apj}
%\bibliography{ws}

%==========================================================================
% Appendix
%==========================================================================
\clearpage
\appendix

%==========================================================================
% Appendix A: Changes Since Paper V
%==========================================================================
\section{Thermal Diffusion}\label{Sec:Thermal Diffusion}
Here we describe the changes from Paper V due to the inclusion of the
thermal conduction term in equation (\ref{eq:Enthalpy Equation}).
The boldface notation refers to specific steps in the algorithm, which are
fully described in Appendix A.4 in Paper V.

Applying the chain rule to the equation of state, $h = h(p_0,T,X)$, we note 
that the temperature gradient can be expressed as
\begin{equation}
\nablab T = \frac{1}{c_p}\nablab h + \sum_k\frac{\xi_k}{c_p}\nablab X_k + \frac{h_p}{c_p}\nablab p_0.
\end{equation}
Whenever we require an explicit computation of the thermal conduction term,
we use this formulation.  In the edge state prediction 
({\bf Steps 4H} and {\bf 8H}), we add an 
explicit contribution of the thermal conduction term to the forcing.
Also, whenever we compute the expansion term $S$, we include the
thermal conduction contribution.  To compute thermodynamic derivatives
of this term, we use $h, X$, and $p_0$ as inputs to the equation of state.  
We account for thermal diffusion in the cell update step by replacing
{\bf Steps 4I} and {\bf 8I} with the semi-implicit approach described
below.

%--------------------------------------------------------------------------
% STEP 4I
%--------------------------------------------------------------------------
\noindent {\bf Step 4I.} {\em Diffuse the enthalpy through a time interval of $\dt$.}

Compute $\kth^{(1)}, c_p^{(1)}$, and $\xi_k^{(1)}$ from $\rho^{(1)},
T^{(1)}$, and $X_k^{(1)}$ as inputs to the equation of state.  We
denote the result for enthalpy in {\bf Step 4H} of Paper V as
$(\rho h)^{(2'),*}$ rather than $(\rho h)^{(2),*}$ to indicate that we
are about to account for thermal diffusion and define 
$\rho^{(2'),\star}=\rho^{(2),\star}$.  The update is given by
\begin{eqnarray}
(\rho h)^{(2),\star} &=& (\rho h)^{(2'),\star} + \frac{\dt}{2}\nablab\cdot\left(\frac{\kth^{(1)}}{c_p^{(1)}}\nablab h^{(2),\star} + \frac{\kth^{(1)}}{c_p^{(1)}}\nablab h^{(1)}\right)\nonumber\\
&&- \frac{\dt}{2}\sum_k\nablab\cdot\left(\frac{\xi_k^{(1)}\kth^{(1)}}{c_p^{(1)}}\nablab X_k^{(2),\star} + \frac{\xi_k^{(1)}\kth^{(1)}}{c_p^{(1)}}\nablab X_k^{(1)}\right)\nonumber\\
&&- \frac{\dt}{2}\nablab\cdot\left(\frac{h_p^{(1)}\kth^{(1)}}{c_p^{(1)}}\nablab p_0^{(2),\star} + \frac{h_p^{(1)}\kth^{(1)}}{c_p^{(1)}}\nablab p_0^{(1)}\right),
\end{eqnarray}
which is numerically implemented as a diffusion equation for $h^{(2),\star}$,
\begin{eqnarray}
\left(\rho^{(2),\star} - \frac{\dt}{2}\nablab\cdot\frac{\kth^{(1)}}{c_p^{(1)}}\nablab\right)h^{(2),\star} &=& (\rho h)^{(2'),\star} + \frac{\dt}{2}\nablab\cdot\frac{\kth^{(1)}}{c_p^{(1)}}\nablab h^{(1)}\nonumber\\
&&- \frac{\dt}{2}\sum_k\nablab\cdot\left(\frac{\xi_k^{(1)}\kth^{(1)}}{c_p^{(1)}}\nablab X_k^{(2),\star} + \frac{\xi_k^{(1)}\kth^{(1)}}{c_p^{(1)}}\nablab X_k^{(1)}\right)\nonumber\\
&&- \frac{\dt}{2}\nablab\cdot\left(\frac{h_p^{(1)}\kth^{(1)}}{c_p^{(1)}}\nablab p_0^{(2),\star} + \frac{h_p^{(1)}\kth^{(1)}}{c_p^{(1)}}\nablab p_0^{(1)}\right),
\end{eqnarray}
Then, update temperature using the equation of state: $T^{(2),\star} = T\left(\rho^{(2),\star}, h^{(2),\star}, X_k^{(2),\star}\right)$.

%--------------------------------------------------------------------------
% STEP 8.1
%--------------------------------------------------------------------------
\noindent {\bf Step 8I.} {\em Diffuse the enthalpy through a time interval of $\dt$.}

Compute $\kth^{(2),\star}, c_p^{(2),\star}$, and $\xi_k^{(2),\star}$,
from $\rho^{(2),\star}, T^{(2),\star}$, and $X_k^{(2),\star}$ as
inputs to the equation of state.  We also denote the result for
enthalpy in {\bf Step 8H} of Paper V as $(\rho h)^{(2')}$ rather
than $(\rho h)^{(2)}$ to indicate that we are about to account for
thermal diffusion and define $\rho^{(2')}=\rho^{(2)}$.  The update is given by
\begin{eqnarray}
(\rho h)^{(2)} &=& (\rho h)^{(2')} + \frac{\dt}{2}\nablab\cdot\left(\frac{\kth^{(2),\star}}{c_p^{(2),\star}}\nablab h^{(2)} + \frac{\kth^{(1)}}{c_p^{(1)}}\nablab h^{(1)}\right)\nonumber\\
&&- \frac{\dt}{2}\sum_k\nablab\cdot\left(\frac{\xi_k^{(2),\star}\kth^{(2),\star}}{c_p^{(2),\star}}\nablab X_k^{(2)} + \frac{\xi_k^{(1)}\kth^{(1)}}{c_p^{(1)}}\nablab X_k^{(1)}\right)\nonumber\\
&&- \frac{\dt}{2}\nablab\cdot\left(\frac{h_p^{(2),\star}\kth^{(2),\star}}{c_p^{(2),\star}}\nablab p_0^{(2)} + \frac{h_p^{(1)}\kth^{(1)}}{c_p^{(1)}}\nablab p_0^{(1)}\right),
\end{eqnarray}
which is numerically implemented as a diffusion equation for $h^{(2)}$,
\begin{eqnarray}
\left(\rho^{(2)} - \frac{\dt}{2}\nablab\cdot\frac{\kth^{(2),\star}}{c_p^{(2),\star}}\nablab\right)h^{(2)} &=& (\rho h)^{(2')} + \frac{\dt}{2}\nablab\cdot\frac{\kth^{(1)}}{c_p^{(1)}}\nablab h^{(1)}\nonumber\\
&&- \frac{\dt}{2}\sum_k\nablab\cdot\left(\frac{\xi_k^{(2),\star}\kth^{(2),\star}}{c_p^{(2),\star}}\nablab X_k^{(2)} + \frac{\xi_k^{(1)}\kth^{(1)}}{c_p^{(1)}}\nablab X_k^{(1)}\right)\nonumber\\
&&- \frac{\dt}{2}\nablab\cdot\left(\frac{h_p^{(2),\star}\kth^{(2),\star}}{c_p^{(2),\star}}\nablab p_0^{(2)} + \frac{h_p^{(1)}\kth^{(1)}}{c_p^{(1)}}\nablab p_0^{(1)}\right),
\end{eqnarray}
Then, update the temperature using the equation of state: $T^{(2)} = T\left(\rho^{(2)}, h^{(2)}, X_k^{(2)}\right)$.

\subsection{Diffusion Solver Test}\label{Sec:Diffusion Solver Test}
This problem is designed to test the accuracy of our
implementation of an implicit solver for the diffusion of a two-dimensional
Gaussian enthalpy pulse.  That is, we are only concerned with the
diffusive term in (\ref{eq:Enthalpy Equation}):
\begin{equation}\label{eq:Enthalpy Diffusion}
\frac{\partial (\rho h)}{\partial t} = \nablab \cdot \left(\kth\nablab T\right).
\end{equation}
To easily compare with an analytic solution (see, for example
\cite{SWE_MYRA09} for an analogous example for a
radiation-hydrodynamics code) we assume the thermal conductivity to be
constant: $\kth = 10^7 \text{ erg K cm}^{-1} \text{ s}^{-1}$.  Note
that this does not fully test the predictor-corrector aspect of the
method outlined in Appendix \ref{Sec:Thermal Diffusion} because in
this simplified problem $\kth^{(2),\star}=\kth^{(1)}$.  We also assume
an ideal gas with $X(\text{He}^4) = 0.5, X(\text{C}^{12}) =
X(\text{Fe}^{56}) = 0.25$ and ratio of specific heats $\gamma = 5/3$.
Furthermore, we are not concerned with any hydrodynamic motions so we
keep the density fixed.  We can then express (\ref{eq:Enthalpy
  Diffusion}) in a simpler form:
\begin{equation}
\frac{\partial h}{\partial t} = D\nablab^2 h,
\end{equation}
where $D = \kth/\left(\rho c_p\right)$ is the diffusion coefficient.

Given the initial conditions for the two-dimensional pulse,
\begin{equation}
h(\mathbf{r}, t=t_0) = (h_{\text{p}} - h_0) \times \exp\left(\frac{-\left|\mathbf{r}-\mathbf{r_0}\right|^2}{4Dt_0}\right) + h_0,
\end{equation}
where $h_p$, $h_0$, $\mathbf{r_0}=(x_0,y_0)$, and $t_0$ are the peak
enthalpy, ambient enthalpy, location of the center of the peak, and
time from which the system has evolved respectively, the analytic
solution takes on the form
\begin{equation}\label{eq:Analytic Solution}
h(\mathbf{r}, t) = \left(h_\text{p} -
h_0\right)\left(\frac{t_0}{t+t_0}\right)\exp\left(\frac{-\left|\mathbf{r}-\mathbf{r_0}\right|^2}{4D\left(t+t_0\right)}\right)
+ h_0,
\end{equation}
where $t$ is the evolved time.

We solve this problem on a Cartesian grid of size $4$ cm $\times$ $4$
cm with the following parameters: $h_\text{p} = 10.0 \text{ erg
  g}^{-1}$, $h_0 = 1.0 \text{ erg g}^{-1}$, $\mathbf{r_0}=\left(2.0
\text{ cm},2.0 \text{ cm}\right)$, $t_0 = 0.1 \text{ s}$, and $\rho =
1.0 \text{ g cm}^{-3}$.  For the density and composition used in this
test, we obtain a diffusion coefficient of $D = 0.32 \text{
  cm}^2\text{ s}^{-1}$.  As motivated in Appendix \ref{Sec:Thermal
  Diffusion}, our implicit solve uses a Crank-Nicholson scheme that
is second order accurate in space and time.

\begin{figure*}
\begin{center}
\plotone{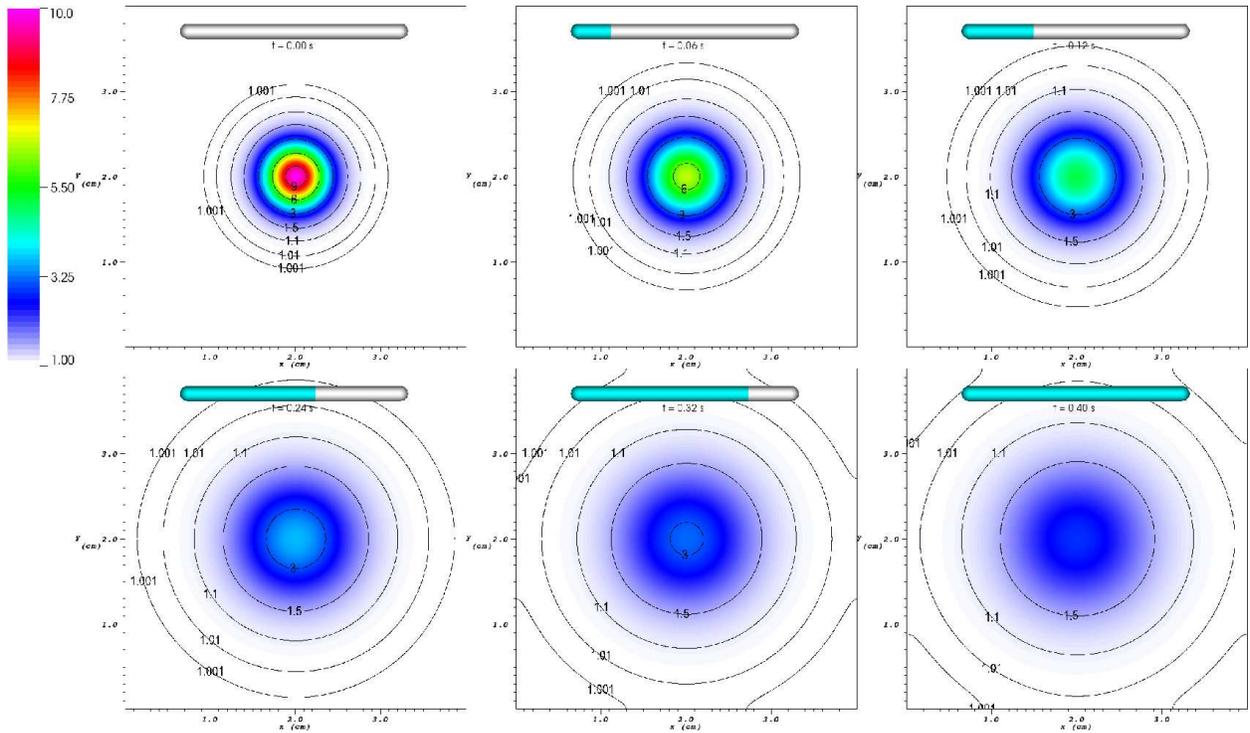}
\end{center}
\caption{\label{Fig:diffusion in 2d} Time evolution of the diffusion
  of a two-dimensional Gaussian pulse of enthalpy as described in the
  text.  The value of time displayed is the evolution time, $t$.  This
  simulation was run with a $1024\times1024$ grid with time step size
  $\Delta t = 0.001$ s.  Excepting edge effects near the domain
  boundary, the numerical solution maintains its axisymmetric form
  about the center of the pulse at $(x,y) = (2.0, 2.0)$.  }
\end{figure*}

\begin{figure*}
\begin{center}
\plotone{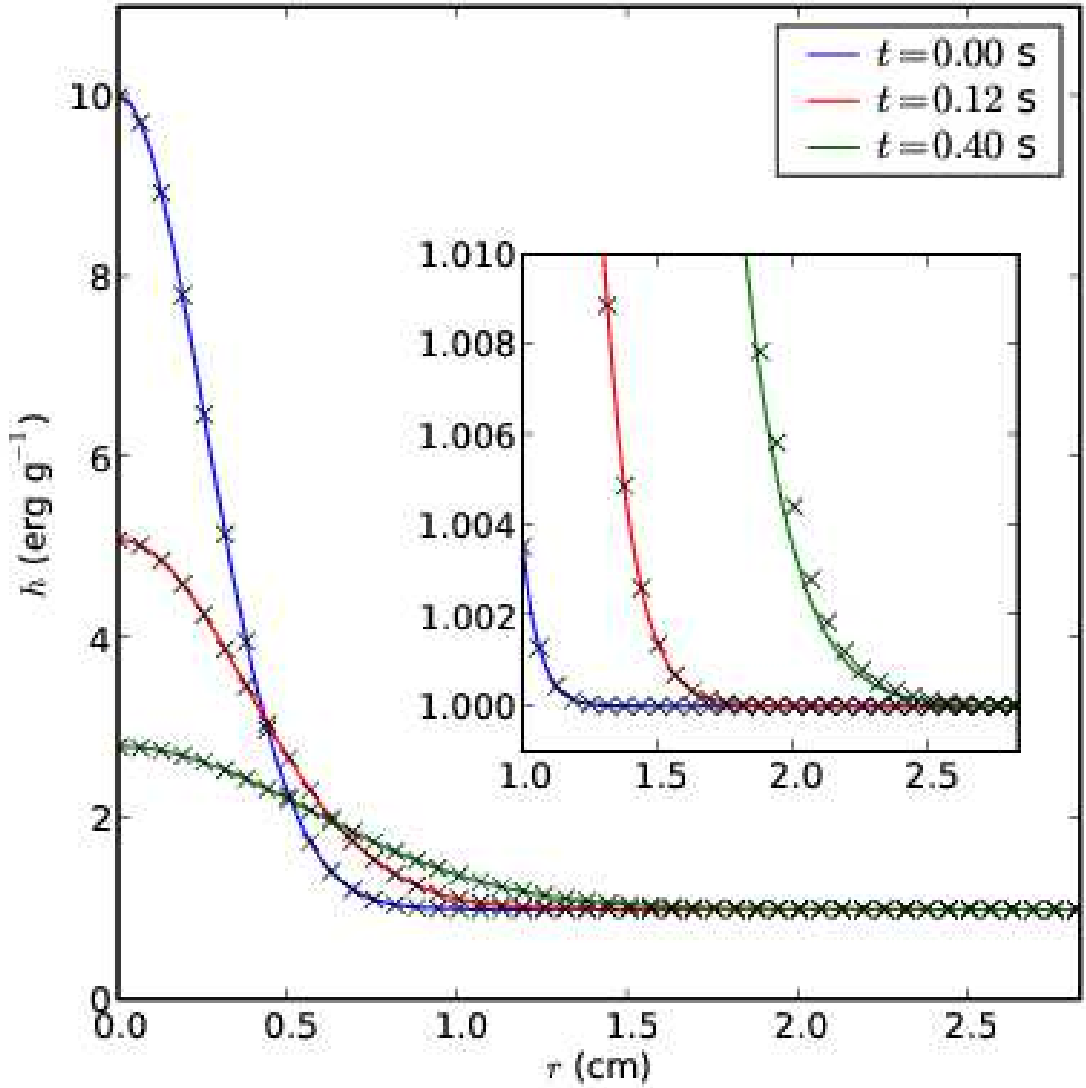}
\end{center}
\caption{\label{Fig:average of enthalpy diffusion} The average of
  enthalpy as a function of radius from the center, $(x,y) =
  (2.0,2.0)$, of a two-dimensional Gaussian pulse.  The X's
  are data from the numerical solution at the shown times.
  The lines represent the analytic solutions as given by
  (\ref{eq:Analytic Solution}).  The numerical solution tracks the
  analytic solution very well except when the pulse has diffused
  enough that it begins to interact with the boundaries of the
  computational domain as seen in the inset plot.}
\end{figure*}

Figure \ref{Fig:diffusion in 2d} shows an example of the initial
enthalpy pulse and its evolution through $t = 0.4$ s on a
$1024\times1024$ grid with fixed time step $\Delta t = 10^{-3}$ s.
Note that as the pulse expands it begins to interact with the edges of
the computational domain and the symmetry of the Gaussian peak is
broken.  Figure \ref{Fig:average of enthalpy diffusion} shows the
computed average enthalpy as a function of radius (X's) compared to
the analytic solution (lines) for the same test problem shown in Figure
\ref{Fig:diffusion in 2d}.  Again, excepting boundary effects the numerical
and analytic solutions are well matched.

To check the convergence of the algorithm we ran simulations with
various resolutions and compared the errors.  To measure the error in
the simulation, we use the $L_1$ norm of the difference
between the analytic and numeric solutions normalized to the $L_1$ norm of the analytic solution,  which we define as $\varepsilon$:
\begin{equation}\label{eq:reduced norm}
  \varepsilon^m \equiv
  \frac{||h(r,t^m)-h^m||_{L_1}}{||h(r,t^m)||_{L_1}} = \frac{\sum_{i,j}
    \left|h(r_{i,j},t^m) - h_{i,j}^m\right|}{\sum_{i,j}
    |h(r_{i,j},t^m)|},
\end{equation} 
where $h(r_{i,j},t^m)$ is the analytic solution at $r_{i,j} =
\left((x_i-x_0)^2+(y_j-y_0)\right)^{1/2}$ and time $t^m = m\Delta t$
and $h_{i,j}^m$ is the numeric solution at $(x_i,y_j)$ and time $t^m$.
We further define the convergence rate, $\alpha$, by comparing the
value of $\varepsilon$ at the current resolution to the value of
$\varepsilon$ at a finer resolution simulation:
\begin{equation}\label{eq:alpha}
  \alpha \equiv
  \log_2\left(\frac{\varepsilon}{\left[\varepsilon\right]_\text{finer}}\right).
\end{equation}
For our comparisons, we take ``finer'' to mean a simulation with twice
the resolution; to compare the simulations at the same physical time,
the finer simulation must have evolved through twice the number of
time steps as the coarser simulation.  If our algorithm truly is second
order accurate in space and time then $\alpha$ should equal $2$.
Table \ref{table:conv} shows the values of $\varepsilon$ and the
convergence rate for various resolutions at $t = 0.08$ s; for
$\alpha$, the norm in the current column is compared to the norm of
the finer resolution simulation in the column to its right.  Our
values of $\alpha$ agree very well with the expected value.

\begin{table}
\begin{center}\caption{\label{table:conv} Reduced $L_1$ norms and convergence rate for the diffusion test problem at $t=0.08$ s.\newline}
\begin{tabular}{ccccc}
  \tableline \tableline 
  & $128\times 128$ Error & $256\times 256$ Error & $512\times 512$ Error 
  & $1024\times 1024$ Error \\ 
  & $\Delta t = 0.008$ s & $\Delta t = 0.004$ s & $\Delta t = 0.002$ s 
  & $\Delta t = 0.001$ s 
  \\ 
  \tableline 
  $\varepsilon$ & $8.64\times10^{-5}$ & $2.16\times10^{-5}$ 
  & $5.39\times10^{-6}$ & $1.35\times10^{-6}$ 
  \\ $\alpha$ & $2.0012$ & $1.9999$ & $1.9988$ &---\\ 
  \tableline
\end{tabular}
\end{center}
\end{table}

\clearpage

%%%%%%%%%%%
% FIGURES %
%%%%%%%%%%%

\clearpage

\clearpage

\end{document}